\documentclass{ws-ijmpd}
\usepackage[super,compress]{cite}
\usepackage{color}
\usepackage{ulem}
\usepackage{graphicx}

\begin{document}

\markboth{Hernandez-Almada, Garc\'{\i}a-Aspeitia}{Multistate scalar field dark matter and its correlation with galactic properties}

%
\catchline{}{}{}{}{}
%

\title{Multistate scalar field dark matter and its correlation with galactic properties}

\author{A. Hern\'andez-Almada}

\address{Facultad de Ingenier\'ia, Universidad Aut\'onoma de Quer\'etaro, Centro Universitario Cerro de las Campanas, 76010, Santiago de Quer\'etaro, M\'exico.\\
ahalmada@uaq.mx}

\author{Miguel A. Garc\'{\i}a-Aspeitia}

\address{Consejo Nacional de Ciencia y Tecnolog\'ia, \\ Av. Insurgentes Sur 1582. Colonia Cr\'edito Constructor, Del. Benito Ju\'arez C.P. 03940, Ciudad de M\'exico, M\'exico.\\
Unidad Acad\'emica de F\'isica, Universidad Aut\'onoma de Zacatecas, Calzada Solidaridad esquina con Paseo a la Bufa S/N C.P. 98060, Zacatecas, M\'exico. \\
aspeitia@fisica.uaz.edu.mx}

\maketitle

\begin{history}
\received{Day Month Year}
\revised{Day Month Year}
\end{history}

\begin{abstract}
In this paper, we search for correlations between the intrinsic properties of galaxies and the Bose-Einstein condensate (BEC) under a scalar field dark matter (SFDM) at temperature of condensation greater than zero. According to  this paradigm the BEC is distributed in several states. Based on the galactic rotation curves collected in SPARC dataset, we observe that SFDM parameters present a weak correlation with most of the galaxy properties, having only a correlation with those related to neutral hydrogen emissions. In addition, we found evidence to support of self-interaction between the different BEC states, proposing that in future studies must be considered crossed terms in SFDM equations. Finally, we find a null correlation with galaxy distances giving support to non-hierarchy of SFDM formation.
\end{abstract}


\ccode{PACS numbers: 98.62.-g, 98.90.+s,95.35.+d }


\section{Introduction}


In $\Lambda$ cold dark matter paradigm ($\Lambda$CDM), the dark matter (DM) and dark energy (DE)\footnote{In this particular case, DE is modeled by a cosmological constant (CC).} are the dominant components, being the 96\% of our Universe, according to recent results from several observations \cite{Percival:2009xn,Komatsu:2011,Sharov:2014voa,Planck:2015XIII,Planck:2015XIV,Riess:2016}. However, different evidence shows that DE only influences at cosmological scales, causing an accelerated rate of expansion in the Universe, while DM affects both, galactic and cosmological scales, being fundamental for the growth of structure.

Centering our attention on dark matter, some famous papers by Zwicky \cite{Zwicky1,Zwicky2,Zwicky3}, mark the first evidence\footnote{There is a discussion that Poincare was the first to notice this effect in galactic clusters (see \cite{Bertone:2016nfn} for details).} of a large velocity dispersion exhibited in Coma cluster, suggesting a lack of matter in galactic neighborhoods. Another important step in the study of DM comes from galactic rotation curves, being the arduous work done by different people \cite{Bertone:2016nfn}, culminated in the revolutionary studies performed by Rubin and Ford \cite{RF}. In this last case, it is shown a strong and undebatable evidence of missing matter in galaxies, what is later known as the problem of DM.

Until now, the preferred candidate (see \cite{Bertone:2016nfn} for other candidates) comes from some of the supersymmetric massive particles (SUSY), which meet the necessary characteristics to be the elusive dark matter and being the main ingredient in the $\Lambda$CDM model.
However the inability of a direct detection of a SUSY particles and several problems in the $\Lambda$CDM model, like the overabundance of subhalos predicted \cite{0004-637X-830-2-83} and the \emph{core}-\emph{cusp} problem in density profiles \cite{Navarro1997,Subramanian2000,Salucci}, has led the scientific community to propose radicals new models to solve the DM problem. 

Among the plethora of ideas, some of the most competitive models come from scalar fields, like axions \cite{Lee:1995af,Barranco:2010ib} or the ultralight SFDM \cite{Matos:2000ki,UrenaLopez:2000aj,Matos:2000ss,Matos:2001pz,Rodriguez-Meza:2012,RodriguezMeza/Others:2001,RodriguezMeza/CervantesCota:2004,Rodriguez-Meza:2012b}. The axions are theorized as a solution to the strong CP problem in quantum chromodynamics with massed of $m\sim10^{-7}\rm eV$, meanwhile the SFDM consists in auto-interacting real scalar field whose particles with mass of $m\sim10^{-22}\rm eV$ (fitted by cosmological observations \cite{Matos:2000ki}) condensates forming a BEC in the very early stages of the Universe evolution \cite{Matos:2000ki,UrenaLopez:2000aj,Matos:2000ss,Matos:2001pz,0004-637X-763-1-19}. 

Other important characteristics of SFDM is that if the BEC is at temperature of condensation $T_{\rm crit}=0$ \cite{0004-637X-763-1-19}, there will be only one state of condensation. In contrast, if it is  considered a non-zero temperature of condensation $T_{\rm crit}\sim m^{-5/3}\sim\rm TeV$, the BEC will be arranged in multiple states (multistate). It is notorious that the SFDM model has the characteristic that the haloes of galaxies do not form hierarchically, being the galaxies very similar between them. Indeed, it is expected well-formed galaxy haloes at high redshifts than in the $\Lambda$CDM model. In addition, it is expected no cusp in the center of galaxies and a reduced number of small satellites \cite{Barkana}. Some other SFDM benefits are highlighted by Robles and Matos \cite{0004-637X-763-1-19} as follows: The concordance with the cosmological densities, the fit with the rotational curves of large galaxies and LSB galaxies, the compatibility with observations of the amount of substructure, are among the most important. 

On the other hand, the ultra-light mass associated with the scalar field is its main problem, due to the inability to detect the particle directly at least with current technology. Finally, one interesting way of quantifying the interactions with baryons and DM is through the mass discrepancy-acceleration relation (MDAR) \cite{Ludlow:2016qzh} which it is also applied to the SFDM case \cite{Urena-Lopez:2017tob} in order to quantify the interactions of baryons with DM or SFDM.

Motivated by these ideas, this paper is devoted to study the multistate case of SFDM (BEC) at galactic scale. Accordingly, our aim is to search for correlations between these SFDM multistate and the properties of the galaxies such as luminosity, radius, mass and other primordial characteristics of the galaxies. 
In this paper we start, defining the correlations as a linear dependence between two variables as it is later shown, in order to interpret, analyze and find coincidences between different galaxy properties and the SFDM model (see details in Sec. \ref{data}). 

In this tenor, the study is carried out by performing a Markov Chain Monte Carlo (MCMC) Bayesian analysis with 175 galaxies collected in SPARC catalogue \cite{SPARC:2016}. We focus mainly in the size of the halo, the total halo mass, the central density of the ground state, the central density of the excited states and the number of states, which are in the SPARC sample. For instance, notice that luminosity always have strong correlation with SFDM and it is the most studied characteristic of galaxies. Hence and as a complement, we search for evidence of other kind of correlations with the purpose of not having overlooked some critical details of SFDM.

We organize the paper as follows: In Sec. \ref{SFDM1} we discuss a brief introduction to SFDM model, based on the most recent research in this vein. The previous section is divided into the following subsections, in order to detail our study.
\begin{enumerate}[(a)]
\item[a)] In Sec. \ref{SFDM} we present the density profile associated to SFDM and its corresponding rotation velocity, together with other important definitions like logarithmic slope and the logarithmic rotation curve.
\item[b)] Also, in Sec. \ref{data} we present the data sample and fits by using the MCMC method.
\end{enumerate}
Our results and discussions are presented in Sec. \ref{Res} and finally the conclusions and outlooks are shown in Sec. \ref{Con}.

\section{Revisiting SFDM in galaxy dynamics} \label{SFDM1}

The SFDM model was born with efforts produced by Refs. \cite{Sin} and \cite{Matos:2000ki}, creating the foundations of this novel paradigm to explain DM. However, these foundations were primarily cosmological and later applied to studies of galactic dynamics as shown in Refs. \cite{Matos:2008ag,Nunez:2010ug,Robles:2012uy,Lora:2011yc,0004-637X-763-1-19,Robles:2014ysa,1475-7516-2015-12-025}.

We start the revision, studying the Newtonian approach. In this case, the equation that must be solved can be written in the form \cite{0004-637X-763-1-19}:
\begin{equation}
\delta\ddot{\Phi}-\nabla^2\delta\Phi+\frac{\lambda}{2}(T_C^2-T^2_{\Phi})\delta\Phi=0, \label{1}
\end{equation}
where $T_C$ is related with the critical temperature, $\lambda$ is the Compton wave longitude, $T_{\Phi}=T(t_{form})$, $t_{form}$ is the time in which the halo forms and $\delta$ corresponds to the perturbations to Einstein's field equations in the form $\delta G^{\mu}_{\nu}=8\pi G\delta T^{\mu}_{\nu}$, from where is deduced Eq. \eqref{1} (see \cite{0004-637X-763-1-19} for details). Therefore, the \emph{ansatz} proposed for this case will be
\begin{equation}
\delta\Phi=\sum_j\delta\Phi_0^j\frac{\sin(k_jr)}{k_jr}\cos(\omega_jt), \label{delta}
\end{equation}
where 
\begin{equation}
\omega^2_j=k^2_j+\frac{\lambda}{2}(T^2_C-T^2_{\Phi}),
\end{equation}
being $k_j=j\pi/R$ with $j=1,2,3\dots$ is a parameter related to the radius of the SFDM halo $R$ that considered common in all the multistates and also $\delta\Phi$ is the amplitude of the perturbation. Hence, with Eq. \eqref{delta} and the equation of number density, it is possible to find:
\begin{equation}\label{Gauss}
\rho(r)_{\rm SFDM}^j=\rho_0^j\left[\frac{\sin (k_j r)}{k_j r}\right]^2 \,, 
\end{equation}
where $\rho_0^j = \rho_{{\rm SFDM}}^j(0)$ is the density at $r=0$ of the $j$th excited state, thus $\rho(r)^{tot}_{SFDM}=\sum_j\rho(r)^j_{SFDM}$, it is the sum of all components. In addition we have $T_{\Phi}^2=T^2_C-4\Phi^2_0$ and $\Phi^2_0=\Phi^2_{min}$. 

An important point to notice is that we expect that the sum of the states in Eq. \eqref{delta} is also a solution of Eq. \eqref{1}, because the superposition principle is maintained. In addition and as a first approximation, crossed terms are neglected in this study, following the same procedure of previous authors \cite{0004-637X-763-1-19}.

\subsection{SFDM rotation velocity and some important characteristics} \label{SFDM}

From here, it is possible to present the velocity rotation of galaxies associated with the density profile shown in 
Eq. \eqref{Gauss}. In this vein, it is well known that the rotation velocity at Newtonian limit is related to the 
effective 
potential and is given by
\begin{equation} \label{rotvel}
V^2(r)=r\left\vert \frac{d\Phi(r)}{dr}\right\vert=\frac{G\mathcal{M}(r)}{r}\,, 
\end{equation}
where $\mathcal{M}(r)$ is the total mass which describes the galactic dynamics and can be written for SFDM as:
\begin{equation}\label{mass}
\mathcal{M}^j_{\rm SFDM}(r)=  \frac{2\pi  \rho_0^jr}{k_j^2} \left[1-\frac{\sin (2k_j r)}{2k_j r}\right] \,.
\end{equation}
Using Eq. \eqref{rotvel} and \eqref{mass} we obtain the rotation velocity in the form:
\begin{equation}\label{velrotnon}
V^{j}_{\rm SFDM}(r)= \sqrt{\frac{2\pi G \rho_0^j}{k_j^2} \left[1-\frac{\sin (2k_j r)}{2k_j r}\right]} \,,
\end{equation}
where in both cases are for the $j$th excited state.
In addition, the 
maximum in the rotation curve can be obtained with the following equation
\begin{equation}
\frac{\cos(2k_jr_{max})}{2k_j^2r_{max}}\left[\frac{\tan(2k_jr_{max})}{2k_jr_{max}}-1\right]=0,
\end{equation}
here $r_{max}$ determines the first local maximum of the rotation curve velocity  \cite{0004-637X-763-1-19}.

Therefore, we consider $V_{\rm total}$ as the sum of the gas $V_{\rm gas}$, disk $V_{\rm  disk}$, bulge $V_{\rm bulge}$, and SFDM (halo) $V_{\rm SFDM}$ components. In other 
words,
\begin{equation} \label{eq:Vtotal}
V_{\rm total}^2=V_{\rm gas}^2 + \Upsilon_{\rm d}V_{\rm disk}^2 + \Upsilon_{\rm b}V_{\rm bulge}^2 + V_{\rm SFDM}^2\,,
\end{equation}
where $\Upsilon_{\rm d}$ and $\Upsilon_{\rm b}$ are the stellar mass-to-light ratio corresponding to the disk and bulge components respectively. In general they are functions of the radius; however in this study is considered to be constant: $\Upsilon_{\rm d}=0.4$ and $\Upsilon_{\rm b}=0.7$ (later we will discuss the consequences of a float $\Upsilon$'s). The latter term in Eq. (\ref{eq:Vtotal}) is described by the multistate of the SFDM density as show Ref. \cite{0004-637X-763-1-19}. Therefore, the general solution in the linear regimen for describing the DM component is given by the superposition of several states of the SFDM.

In addition, it is possible to define the logarithmic slope of a density profile and the rotation curve respectively as \cite{Harko1}
\begin{equation}
\alpha^j = \frac{d(\log\rho^j_{{\rm SFDM}})}{d(\log r)}, \;\; \beta^j = \frac{d(\log V^j_{{\rm SFDM}})}{d(\log r)},  \label{eq:8.3} 
\end{equation}
obtaining for the $j$th state of the SFDM, the following expressions
\begin{subequations}
\begin{eqnarray}
&&\alpha^j = 2(k_j r\cot(k_jr)-1), \\ 
&&\beta^j = -\frac{1}{2}\left[\frac{2k_jr\cot(2k_jr)-1}{2k_jr\csc(2k_jr)-1}\right].  \label{eq:8.4} 
\end{eqnarray}
\end{subequations}
In case that we are considering the total density as the sum of the $N$ states, we would have
\begin{subequations}
\begin{eqnarray}
&&\alpha=\frac{\sum_j^N(\rho_0^j/k_j^2)(k_jr\sin(2k_jr)+\cos(2k_jr)-1)}{\sum_j^N(\rho_0^j/k_j^2)\sin^2(k_jr)}\,,\label{eq:alpha}
\\
&&\beta=-\frac{1}{2}\frac{\sum_j^N(\rho_0^j/k_j^2)[(\sin(2k_jr)/2k_jr)-\cos(2k_jr)]}{\sum_j^N(\rho_0^j/k_j^2)[(\sin(2k_jr)/2k_jr)-1]}\,. \label{eq:beta}
\end{eqnarray}
\end{subequations}
In addition and in similitude with \cite{Robles:2012uy}, the core radius can be defined as the radius at which the core begins, in such a way that its value $R_{{\rm core}}$ is expressed by the equation
\begin{equation}\label{eq:alphaRcore}
\alpha(R_{{\rm core}})=-1.
\end{equation}
Following the above equation, it is possible to infer directly whether if the DM model profile is cored or cuspy.

\subsection{Data sample and fits} \label{data}

In order to study the multistate SDFM profile in rotation curve of galaxies, we use the SPARC sample \cite{SPARC:2016} containing 175 galaxies with different properties. Apart of galaxy rotation curves, the catalogue includes their properties such as distance, luminosity, among others. In Tab. \ref{Tab:PropGalaxy} it is summarized the properties taken into account in this work. As was mentioned before, the focus is to search for correlations between these properties and the fitting parameters of the multistate SFDM model used to describe the DM component of the galaxies. Hence, the correlation coefficient between two variables $X$ and $Y$ is defined as
\begin{equation}\label{eq:corr}
\rho_{XY} = \frac{\mathrm{Cov}(X,Y)}{\sigma_X\sigma_Y},
\end{equation}
where $\mathrm{Cov}(X,Y) = <(X - \mu_X)(Y - \mu_Y)>$ is the covariance coefficient, $\mu_Z = <Z>$ is the expected value of $Z$ and $\sigma_Z^2 = <(Z - \mu_Z)^2>$ is the variance of $Z$.

Our procedure starts by getting the best fit of SFDM model to each galaxy data by using MCMC method, implemented in PyMC package \cite{Patil10pymc:bayesian}, that minimizes the merit-of-function $\log \mathcal{L} \sim \chi^2$ where:
\begin{equation}\label{eq:chi2}
\chi^2(\Theta)=\sum_i^N \left( \frac{V_{\mathrm{obs}}^i-V_{\mathrm{total}}(\Theta)}{dV_{\mathrm{obs}}
^i} \right)^2\,,
\end{equation}
representing with $\Theta$ the free parameters of the model, the observed velocity at the radial distance $r_i$ as $V_{\mathrm{obs}}^i$ and its  
uncertainty as $dV_{\mathrm{obs}}^i$. Notice that the free parameters comes only from the DM component. The corresponding density at $r=0$ per state and the radius $R$, that is the same for all the states. Therefore, using the following study presented in Ref. \cite{Urena/Bernal:2010} to obtain stable state halos, the fit is constrained to the ratio between the mass of any excited state and the ground state mass to be less than $1.3$. 

For each galaxy, we generate $500\,k$ chains with a burn-in of $3\,k$ steps. The  priors considered for all parameters are flat in the intervals
\begin{eqnarray}
0.5<R<200\,\rm kpc, \;\; 10^{4}<\rho_0^i<10^{11}\, \rm M_{\odot}/kpc^3,
\end{eqnarray}
for $R$ and the $i$-state density, respectively.

\begin{figure*}[h]
   \centering
        \label{fig:plot_01}         
        \includegraphics[width=0.45\textwidth]{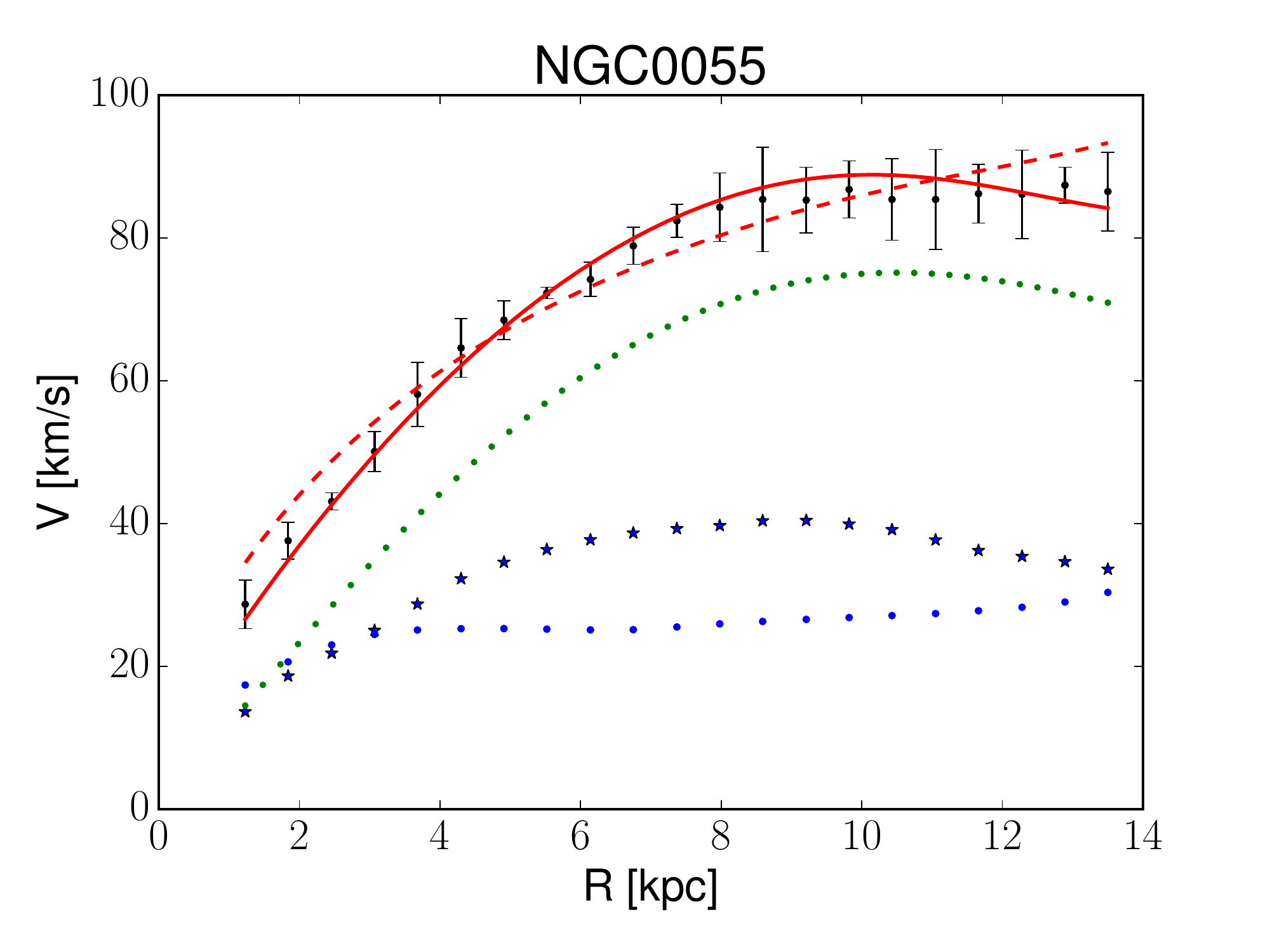}  
        \label{fig:plot_02}         
        \includegraphics[width=0.45\textwidth]{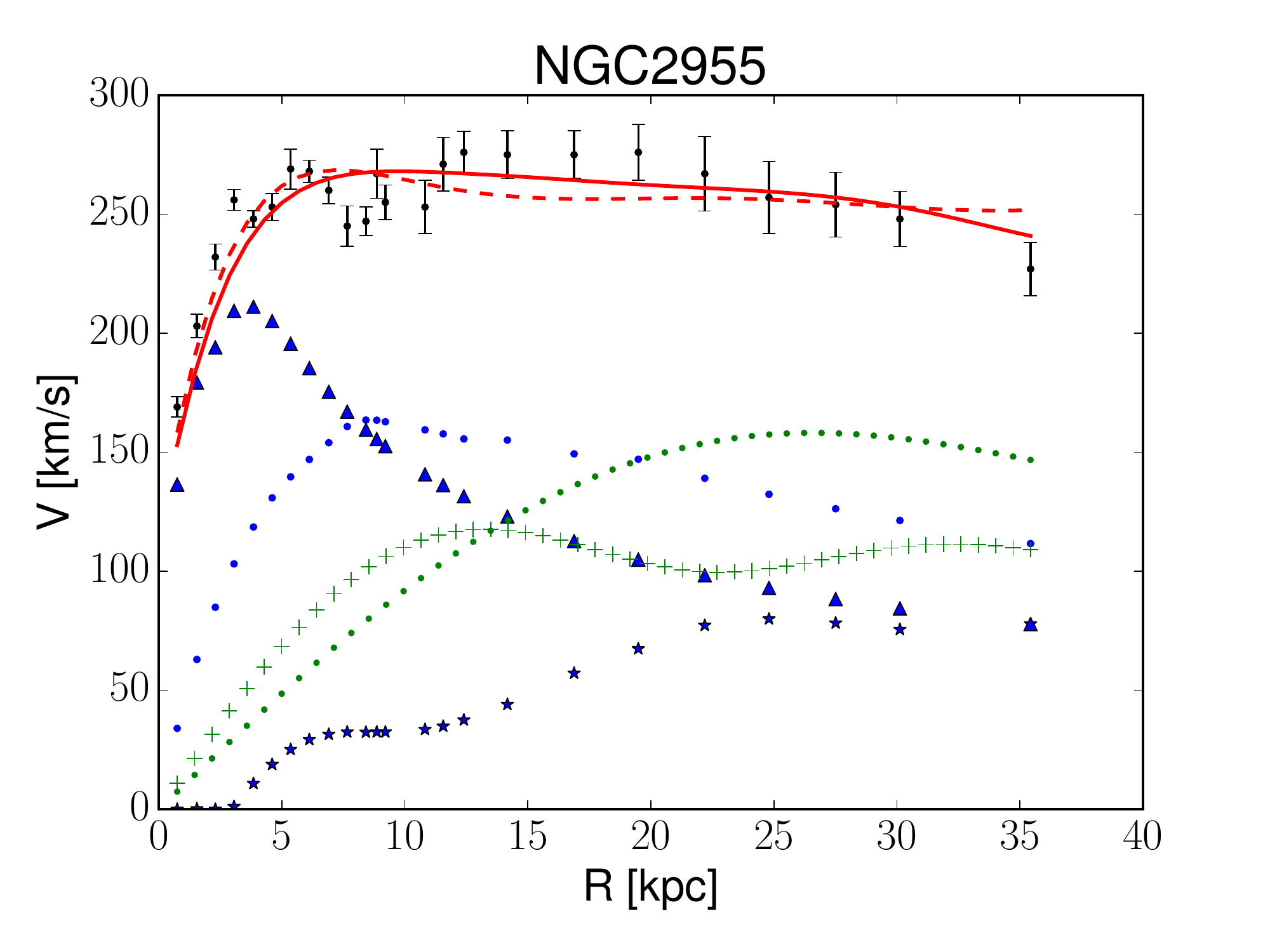} 
   \\
        \label{fig:plot_03}         
        \includegraphics[width=0.45\textwidth]{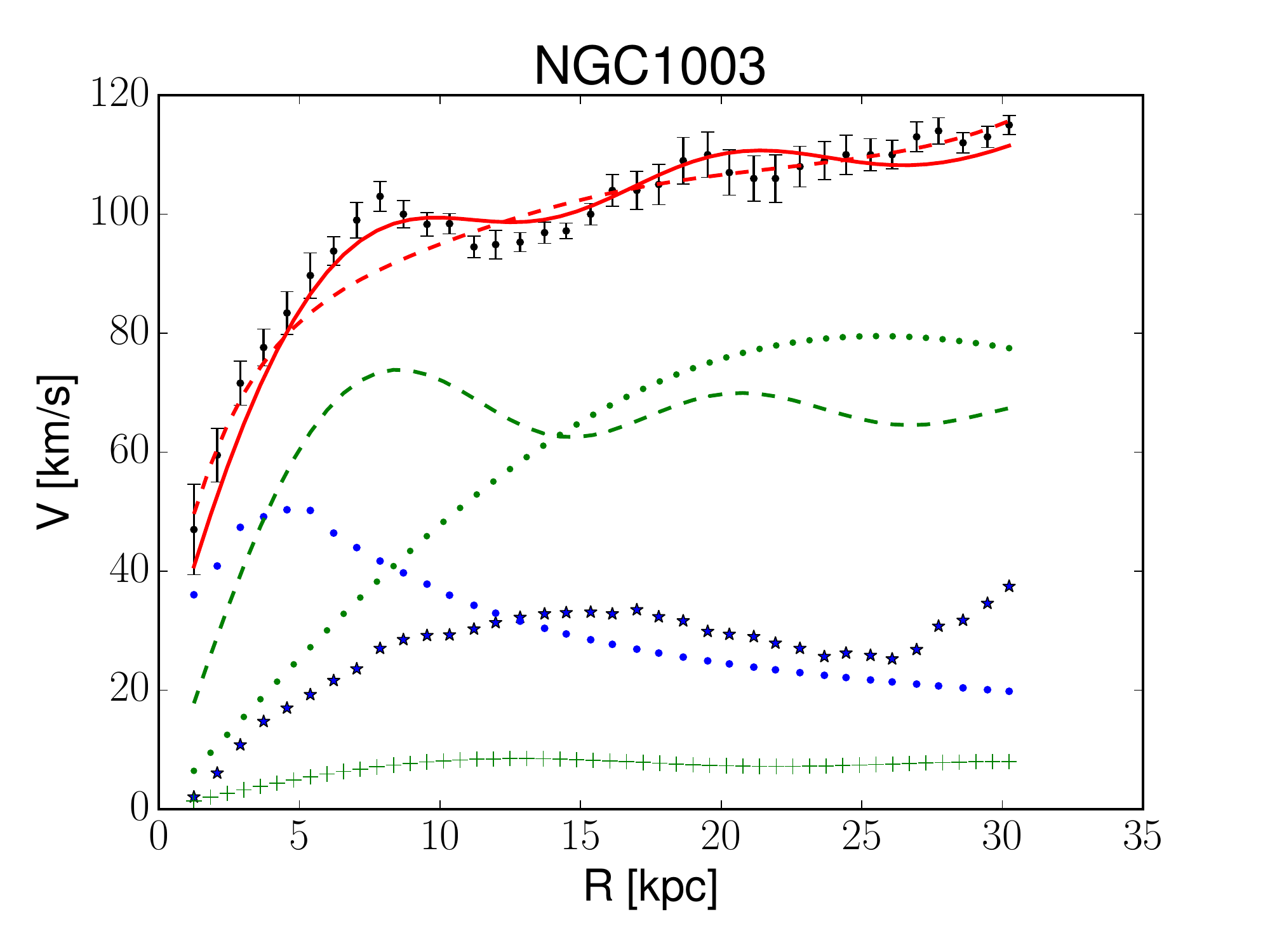}  
        \label{fig:plot_04}         
        \includegraphics[width=0.45\textwidth]{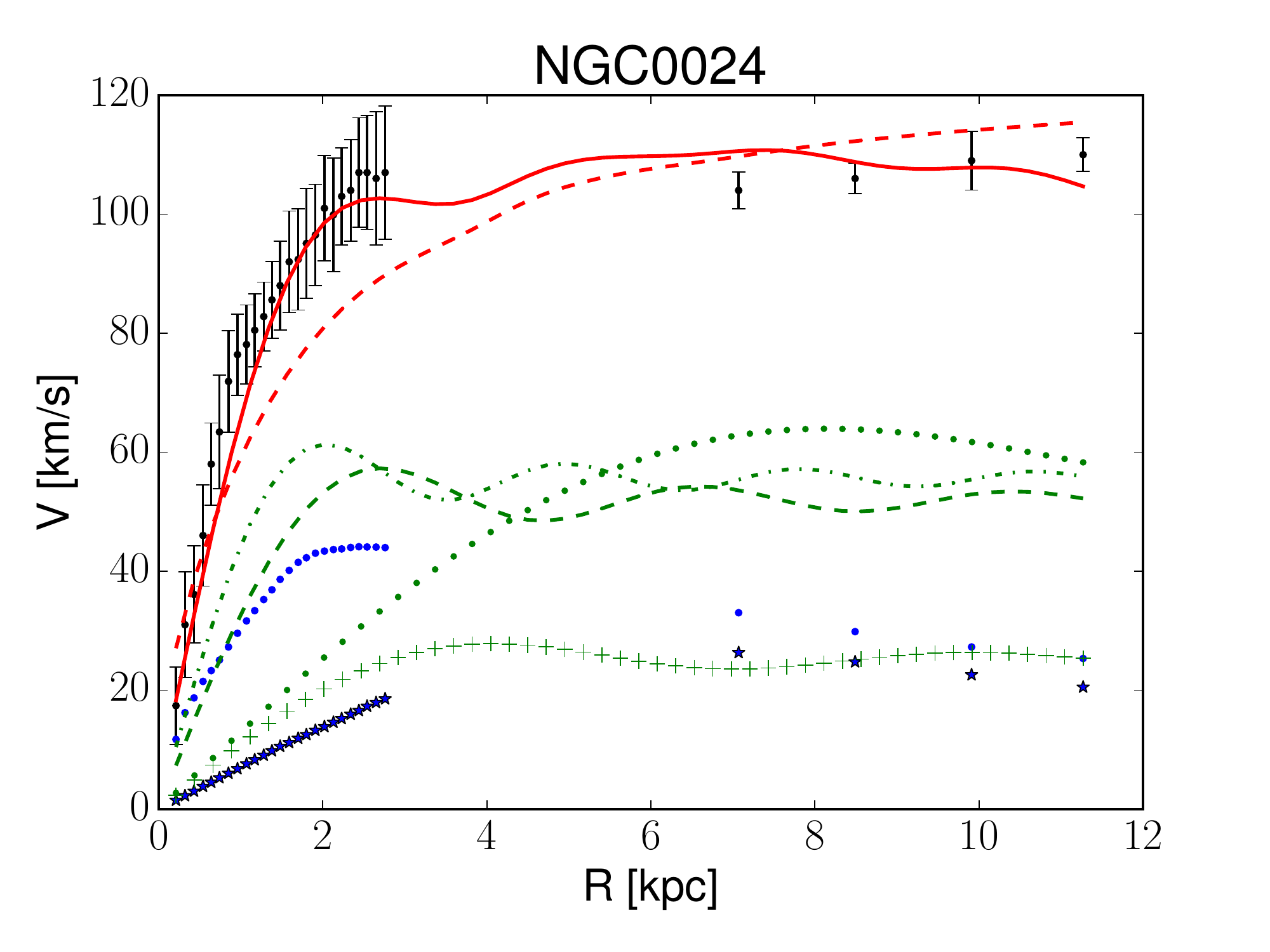} 
   \caption{Best fit of multistate SFDM (solid red line) and NFW \cite{NFW} (dashed red line). On top figures, data are fitted by using one (left) and two (right) states of SFDM and on bottom ones were used three (left) and four (right) components of SFDM. Black points with error bars are the total velocity, blue stars, up-triangles and circles correspond to disk, bulge and gas components respectively. The corresponding components of SFDM are (green color): the ground state with circles, the first excited state with crosses, the second excited state with a dashed line, and third excited state with a dash-dotted line.}
   \label{fig:fits}                
\end{figure*}

The criteria used to get the best DM profile is to select the one that the $\chi^2/{\rm ndf}$ is closest to $1$. Also, we do not consider galaxies with $\chi^2/{\rm ndf}>2$. In Fig. \ref{fig:fits}, we show some examples of fits by using SFDM model and NFW profile (red dashed line). With red solid line is plotted the best fit of multistate SFDM and its components with green color ( the ground state with circles, first excited state with crosses, the second excited state with a dashed line and third excited state with a dash-dotted line according to the case). On top figures are fitted by using one (left) and two (right) states of SFDM and on bottom ones were used three (left) and four (right) components of SFDM.

\begin{table}
\tbl{Galaxy properties from SPARC catalogue. From left to right, we present the properties, meaning and units of the catalogue.}
{\begin{tabular}{@{}ccc@{}} \toprule
Property&Description&Allowance\\ \colrule
$D$& Distance & Mpc \\
$i$&Inclination angle & $(^\circ)$\\
$L_{[3.6]}$&Total luminosity at 3.6$\mu m$&$10^9$L$_{\odot}$\\
$R_{{\rm eff}}$ & Effective radius encompassing half of the total luminosity & $\rm kpc$ \\ 
    $\Sigma_{{\rm eff}}$ & Effective surface brightness & $\rm L_{\odot}/pc^2$ \\ 
    $R_{{\rm d}}$ & Scale length of the stellar disk & $\rm kpc$ \\ 
    $\Sigma_{{\rm d}}$ & Extrapolated central surface brightness of the stellar disk & $\rm L_{\odot}/pc^2$ \\ 
    $M_{{\rm hi}}$ & Total HI mass & $10^9 \rm M_{\odot}$ \\ 
    $R_{{\rm hi}}$ & HI radius where the HI surface density reaches $1 \rm M_{\odot}/pc^2$ & $\rm kpc$ \\ 
    $V_{{\rm f}}$ & Velocity along the flat part of the rotation curve & $\rm km/s$ \\
\botrule
\end{tabular} \label{Tab:PropGalaxy}}
\end{table}

\section{Results} \label{Res}

In order to study the linear correlation between galaxy properties and SFDM fitted parameters, we consider the DM model with up four SFDM states where the ground state is always presented. In this case, we expect that DM should occupy since the innermost states to outermost like occurs with electronic orbital states in atoms. 
Then we focus our attention in four cases: only ground state; ground state plus $1$, $2$ and $3$ excited states, respectively. According to the best fit with $\chi^2/{\rm ndf}$ closest to $1$, we classify the galaxies with respect to number of states. As a rule of thumb to interpret the correlation, we adopt the following convention: A negligible correlation within $[-0.30,0.30]$; a low positive (negative) correlation region as $(0.30, 0.50]$ ($(-0.30, -0.50]$); a moderate positive (negative) region within $(0.50, 0.70]$ ($(-0.50, 0.70]$). Larger positive (lower negative) values than $0.70$ ($-0.70$) correspond to high positive (negative) correlation.

Table \ref{Tab:Corr} shows the correlations between the galactic properties (see Tab. \ref{Tab:PropGalaxy}) and the parameters $R$, total mass $M_{\rm T}$ and $\rho_0^i$, with $i=1,2,3,4$ states, respectively. In the following, we discuss the results obtained for each SFDM parameter. In general, most of the galaxy properties considered in this work are related to the luminosity. For instance, Tully-Fisher relation gives a method to measure the galaxy distance through the luminosity \cite{Tully/Fisher:1977}. Hence, we extend this idea by searching correlation now between galactic properties related mainly with luminosity and halo parameters coming from SFDM model\footnote{For more details about the galactic variables taken into account see Ref. \cite{SPARC:2016}.}.

\begin{figure*}[!ht]
   \centering
        \label{fig:plot_D_vsR}         
        \includegraphics[width=0.48\textwidth]{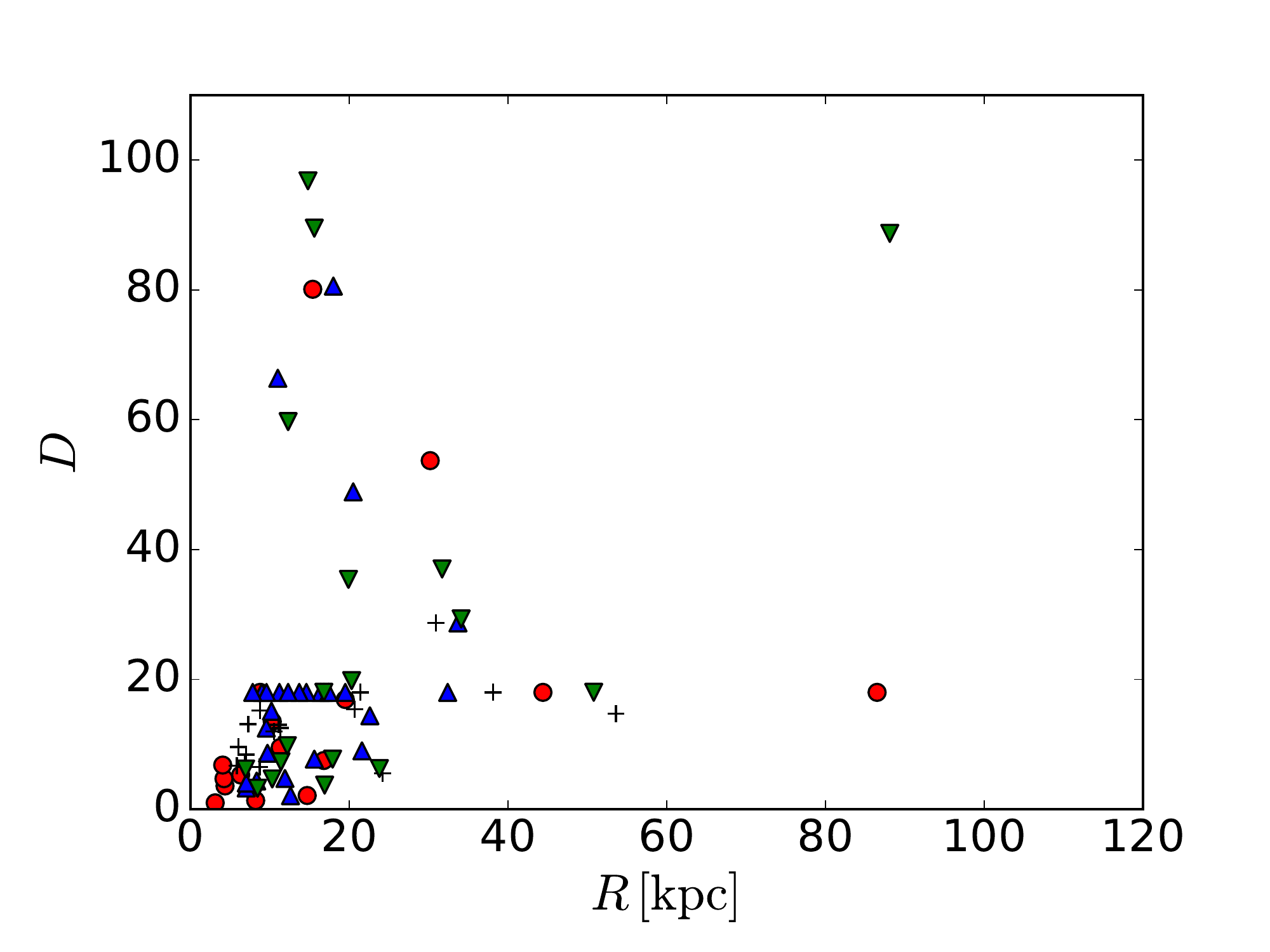}  
        \label{fig:plot_i_vsR}         
        \includegraphics[width=0.48\textwidth]{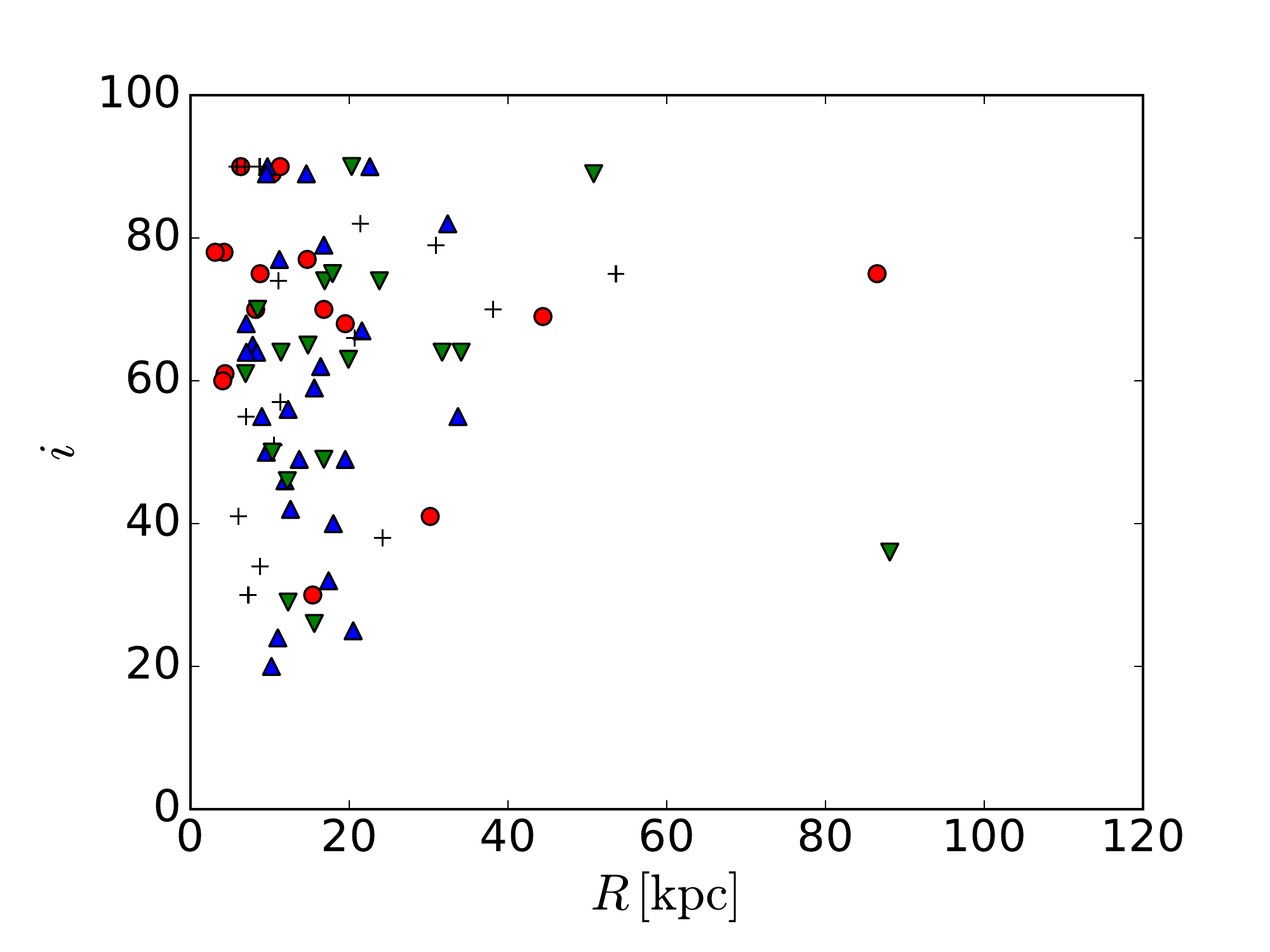} \\
        \label{fig:plot_L_vsR}         
        \includegraphics[width=0.48\textwidth]{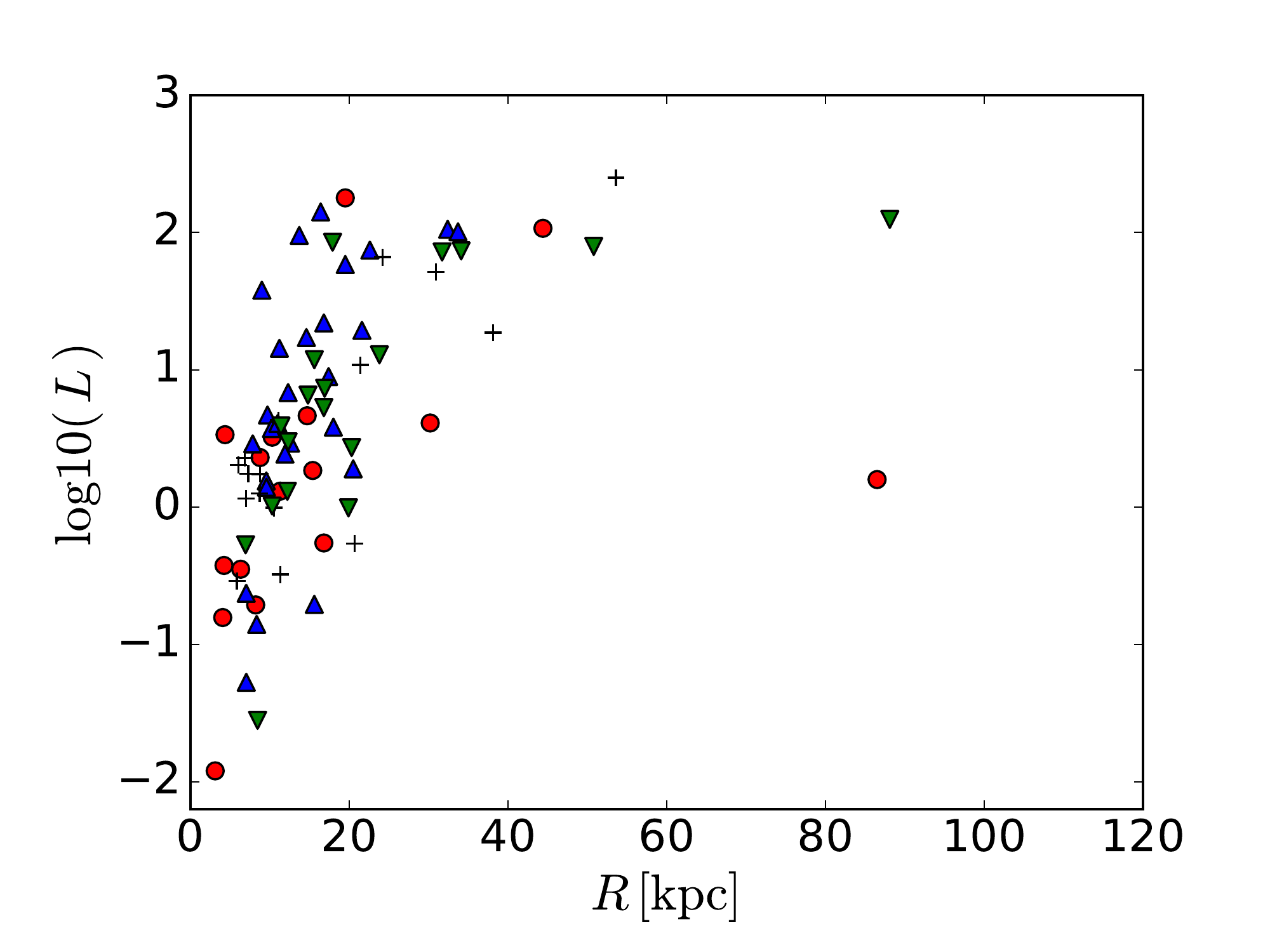}   
        \label{fig:plot_Reff_vsR}         
        \includegraphics[width=0.48\textwidth]{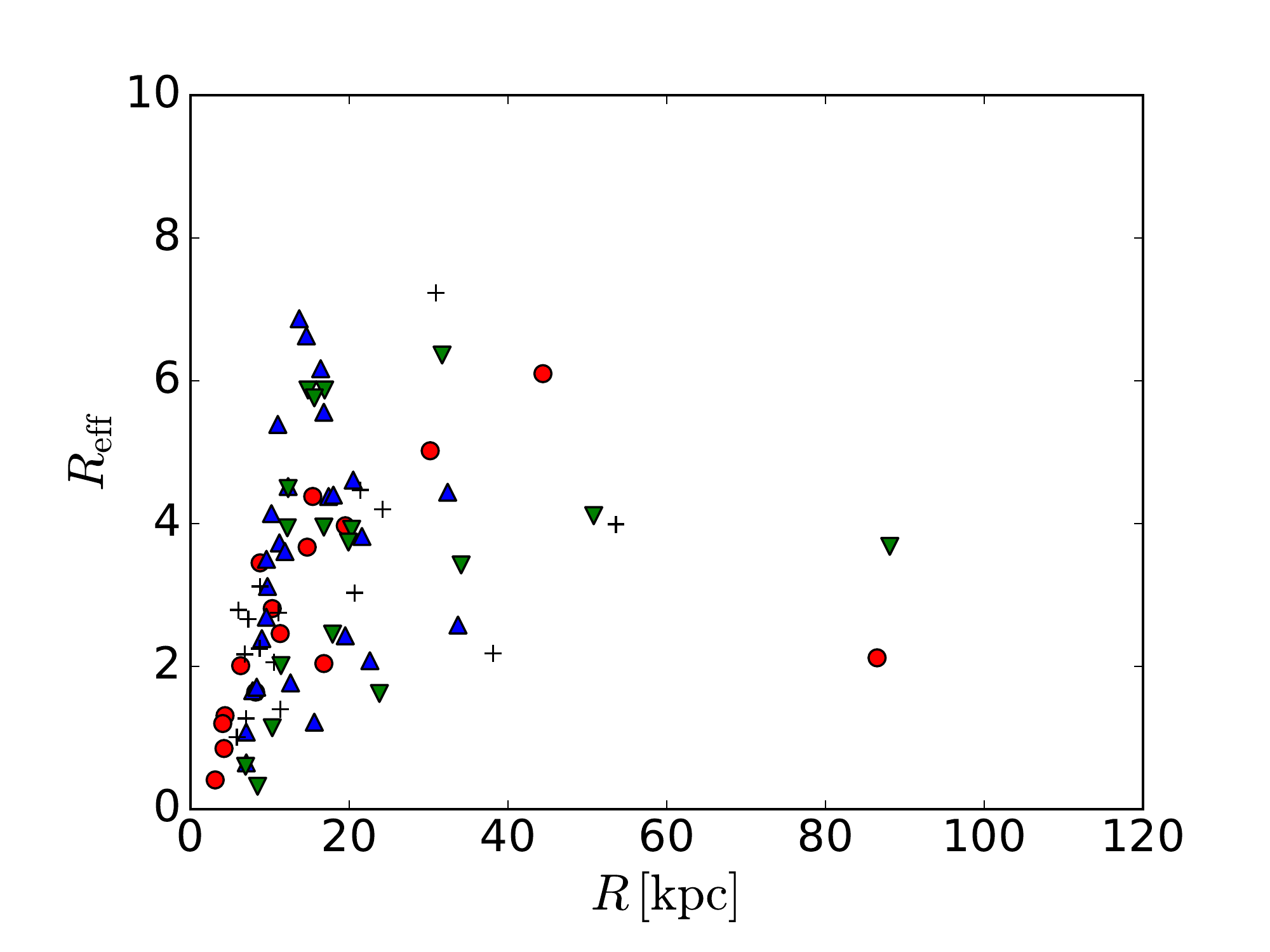} 
   \caption{Scattering of $R$ vs: a) $D$, b) $i$, c) $\log10(L)$, d) $R_{{\rm eff}}$. In all distributions, the number of multistates used to model the DM component is: One (red circles), two (blue up triangles), three (black crosses), and four (green down triangles).}
   \label{fig:VarVsR1}                
\end{figure*}

{\it Halo size} ($R$). The study of the correlation between $R$ and the distance of the galaxies is maybe one of the most interesting correlation. When a null correlation exists,  it may mean that DM halos size does not depend on the distance. According to our results presented in Tab. \ref{Tab:Corr}, we have a weak trend to exist a correlation; however it should come from the fact that far away galaxies are detected when they are luminous enough. In this vein, we can test this hypothesis by removing galaxies larger than $80\,$Mpc (low statistics in this region, see $R$ vs $D$ scattering in  Fig. \ref{fig:VarVsR1}) to re-estimate this correlation, hence we found a correlation of $0.22$ which is consistent with a null correlation.

{\it Total halo mass} $M_{\rm T}$. Due to the high correlation with the size of DM halo, we should have similar trends than $R$. In fact, we give a support of the null correlation between $R$ and $D$.
On the other hand,  we also obtain a negligible correlation with $D$ and $i$ and a low correlation with 
$R_{\rm eff}$ and $\Sigma_{\rm d}$. Finally, we find a moderate correlation with the rest of the galactic parameters.

In addition, we do not find correlation with $i$, $R_{\rm eff}$ and $\Sigma_{\rm d}$. Also, in the region of low positive correlation are $\Sigma_{\rm eff}$, $R_{\rm d}$ and $V_{\rm f}$, and with a moderate positive correlation are $L_{[3.6]}$ and $\Sigma_{\rm eff}$. Finally, we observe a high correlation with the variables related to the neutral hydrogen, $M_{\rm hi}$ and $R_{\rm hi}$. 

\begin{figure*}[h]
   \centering
       \includegraphics[width=0.8\textwidth]{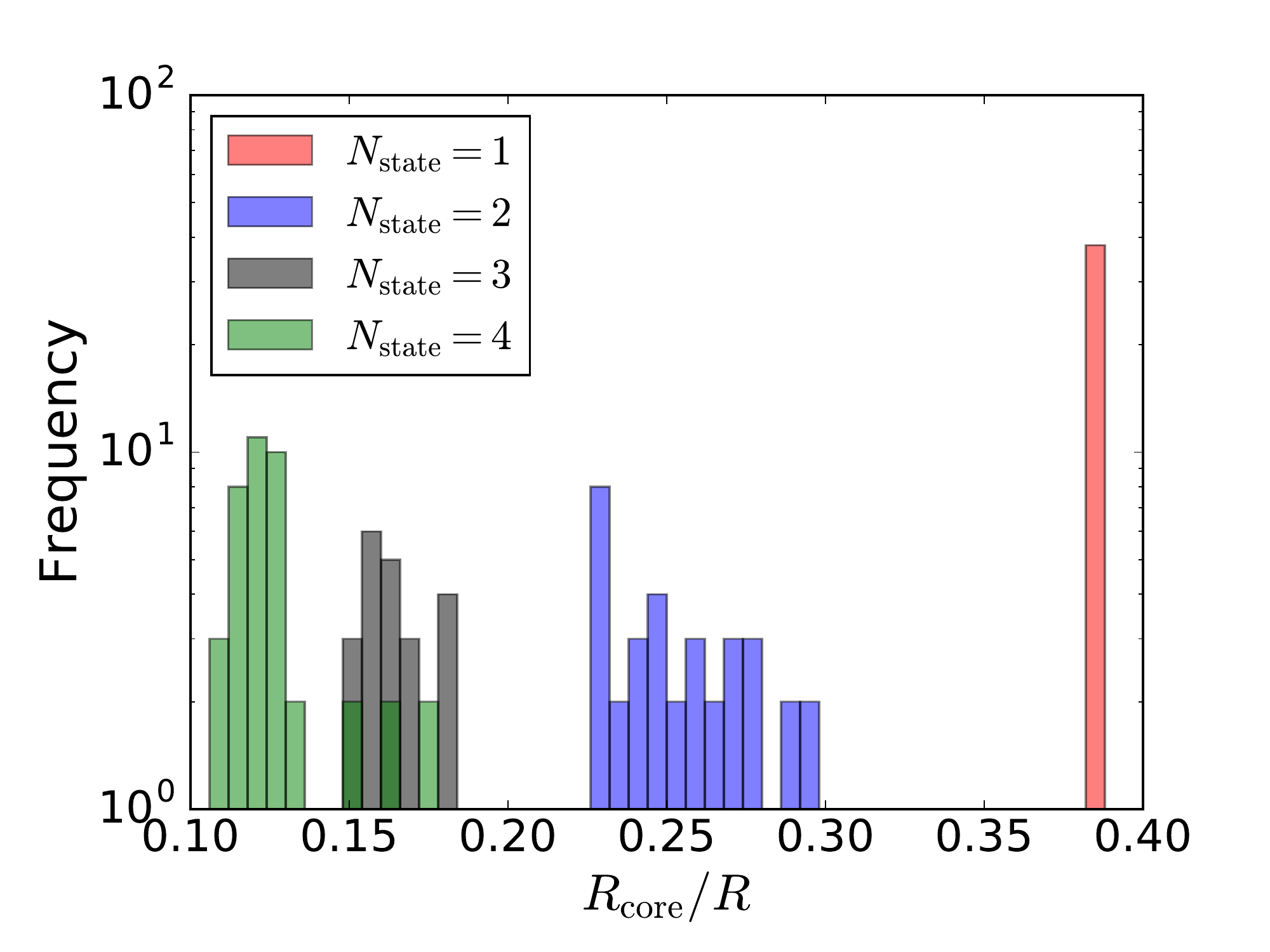} 
   \caption{Distributions of ratio $R_{{\rm core}}/R$ for the four states of the BEC.}
   \label{fig:RcoreOvR}               
\end{figure*}

\begin{figure*}[h]
   \centering
   \includegraphics[width=0.8\textwidth]{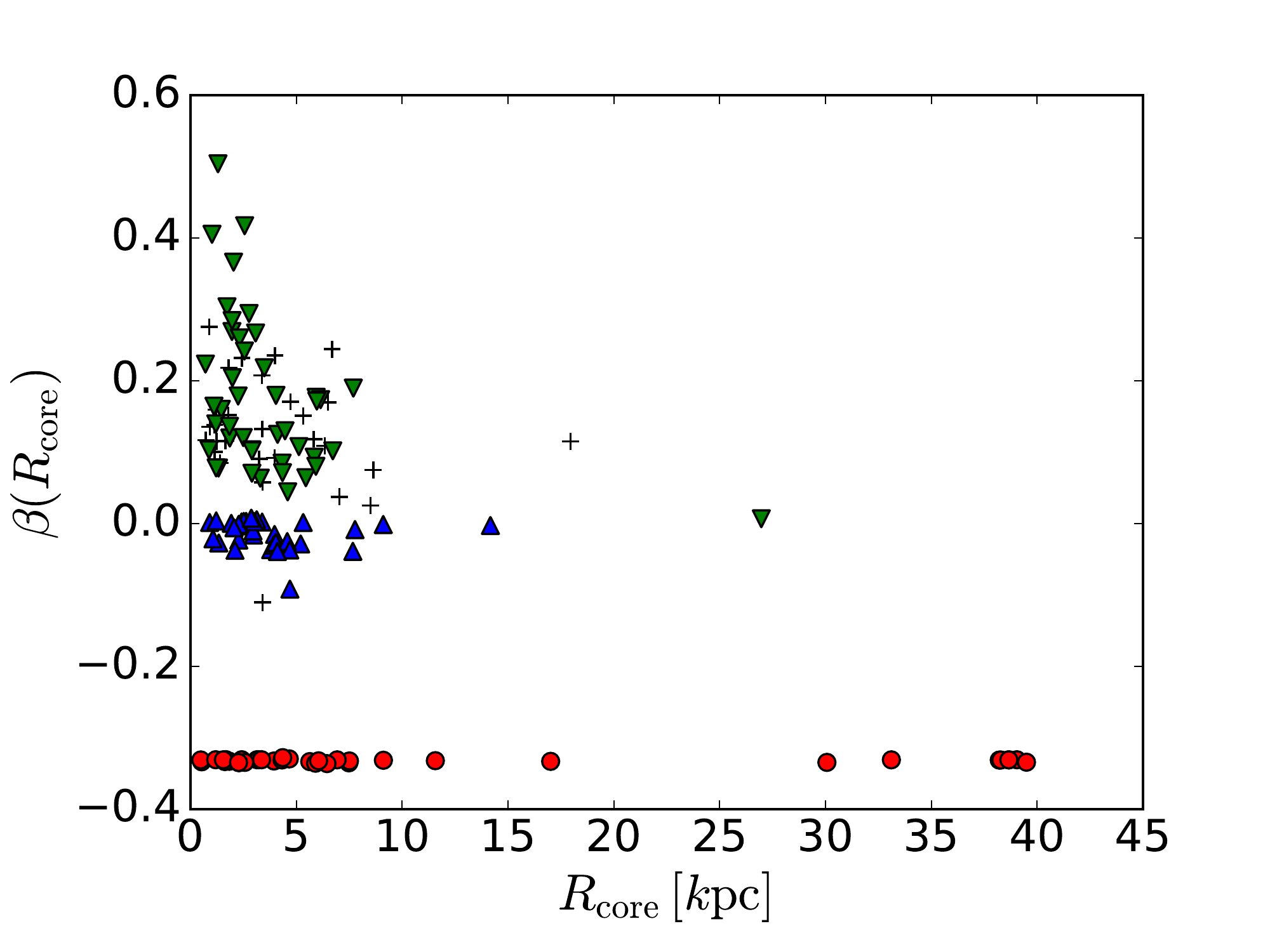}           
   \caption{Scattering distribution of ratio $\beta(R_{{\rm core}})$ vs $R_{{\rm core}}$ for the four states of the BEC. The number of states of DM component is: one (red circles), two (blue up triangles), three (black crosses), and four (green down triangles).}
   \label{fig:BetaRcore}                
\end{figure*}

Furthermore, we estimate the $R_{{\rm core}}$ based on expression (\ref{eq:alphaRcore}) and we found  the median value for the ratio $R_{{\rm core}}/R$ per $N_{\rm state}$ as: 
$0.3800^{+0.0006}_{-0.0001}$, 
$0.25^{+0.03}_{-0.02}$,
$0.17^{+0.03}_{-0.01}$,
$0.12^{+0.04}_{-0.01}$, for $N_{\rm state}=1,\,2,\,3$ and $4$, respectively. Notice that we obtain the largest ratio for the ground state ($N_{\rm state}=1$), and decrease when the number of states increase. This result is shown in Fig. \ref{fig:RcoreOvR}. Also, by using Eq. (\ref{eq:beta}) we obtain the values of $\beta (R_{\rm core})$ per $N_{\rm state}$ as it is presented in Fig. \ref{fig:BetaRcore} as scattering distribution. It is notorious how the base states tend to cluster for values of $\beta(R_{core})<-0.2$, while excited states tends to cluster for $\beta(R_{core})>-0.2$.

{\it Central density of the ground state} ($\rho_0^1$). We find a low negative correlation between $R_{\rm eff}$ and $M_{\rm hi}$, $R_{\rm hi}$, respectively and a null correlation with the other properties. 

It is important to remark, an interesting strong correlation found, is within the densities of the states. Under the hypothesis of uncoupling states in the multistate SFDM model, we observe a considerable correlation of them. Further studies will be needed, taking into account such cross terms in the theory.

{\it Central density of the excited states} ($\rho_0^i$ with $i=2,3,4$). For the first excited state, we observe a null correlation with all galactic properties. For the second one, we have low negative correlation with $L$, $R_{\rm eff}$, $M_{\rm hi}$, $R_{\rm hi}$; a moderate negative correlation with $R_{\rm d}$ and for the rest of the properties a negligible correlation. For the last excited state, the half of the properties considered are not correlated with $\rho_0^4$, while the rest of them are contained in the negative region of low and moderate correlation.  

\begin{figure*}[h]
   \centering
   \includegraphics[width=0.8\textwidth]{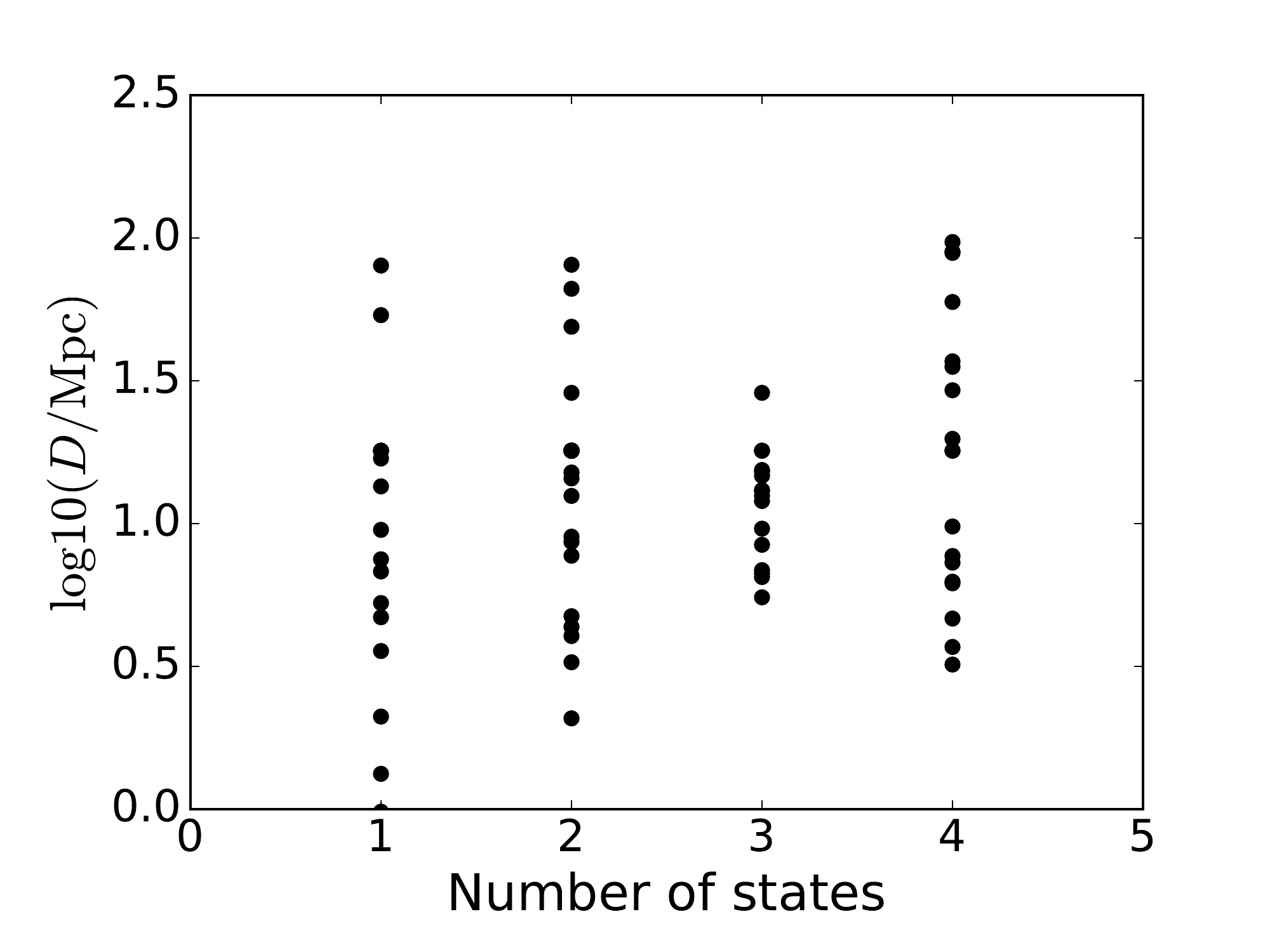}           
 
   \caption{Scattering distribution of $\log_{10}(D)$ vs number of SFDM states.}
   \label{fig:NstVsD}                
\end{figure*}

{\it Number of states}. Figure \ref{fig:NstVsD} shows the distribution of the galaxy distance vs. the number of SDFM states. The distance of the galaxies for $N_{\rm state}=3$ is bounded in the region of $~\,3-32\, \mbox{Mpc}$.  Also, we establish a limit of the galaxy distance to be $>1.23$, $3.47$, $6.17$, $3.89\,$Mpc  for 
$N_{\rm state}=1,2,3,4$ at $95\%$ CL, respectively. It could imply a positive relationship between these two 
variables.

\begin{figure*}[!ht]
   \centering  
        \label{fig:plot_Rd_vsR}         
        \includegraphics[width=0.48\textwidth]{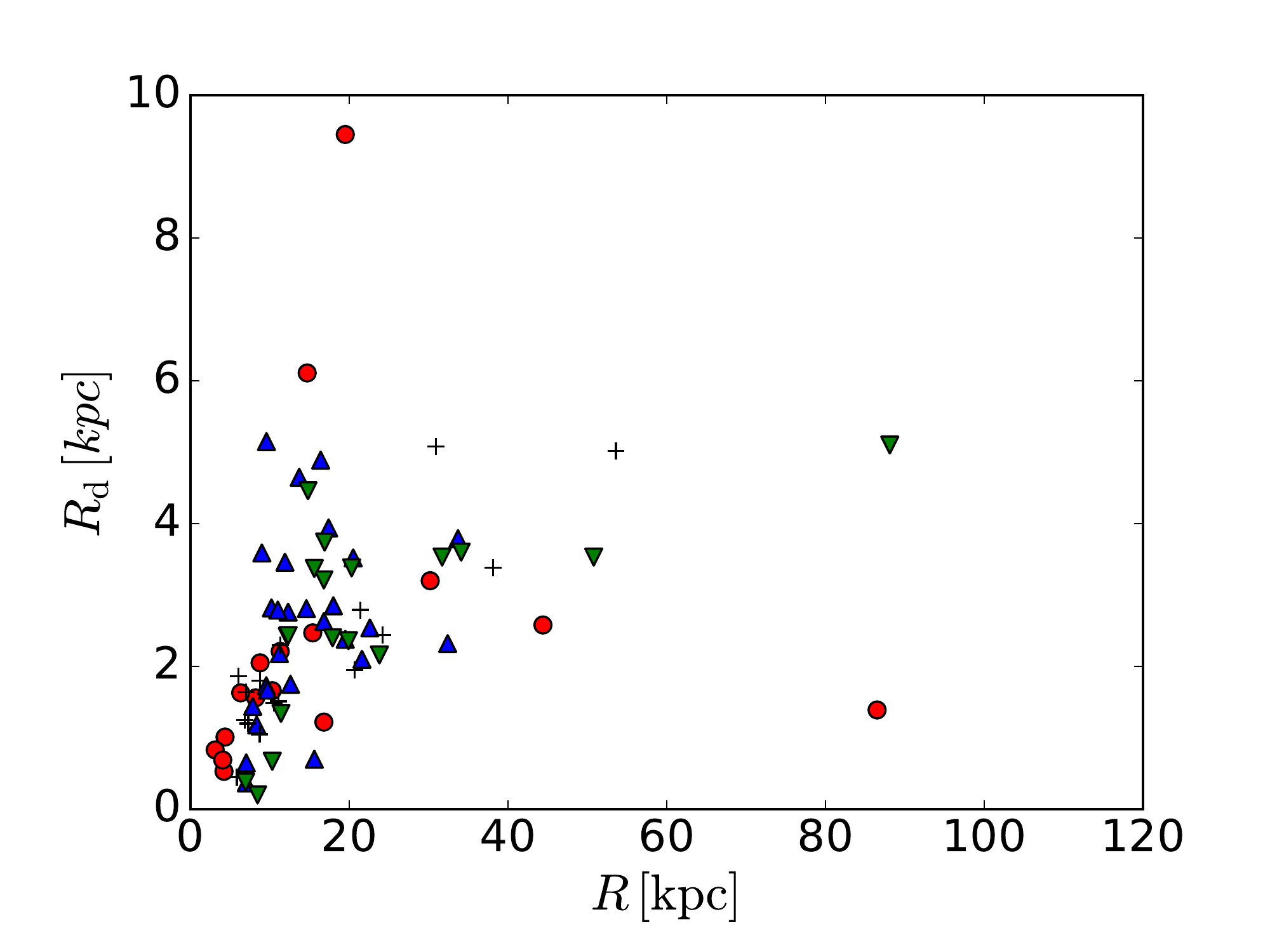}   
        \label{fig:plot_Mhi_vsR}         
        \includegraphics[width=0.48\textwidth]{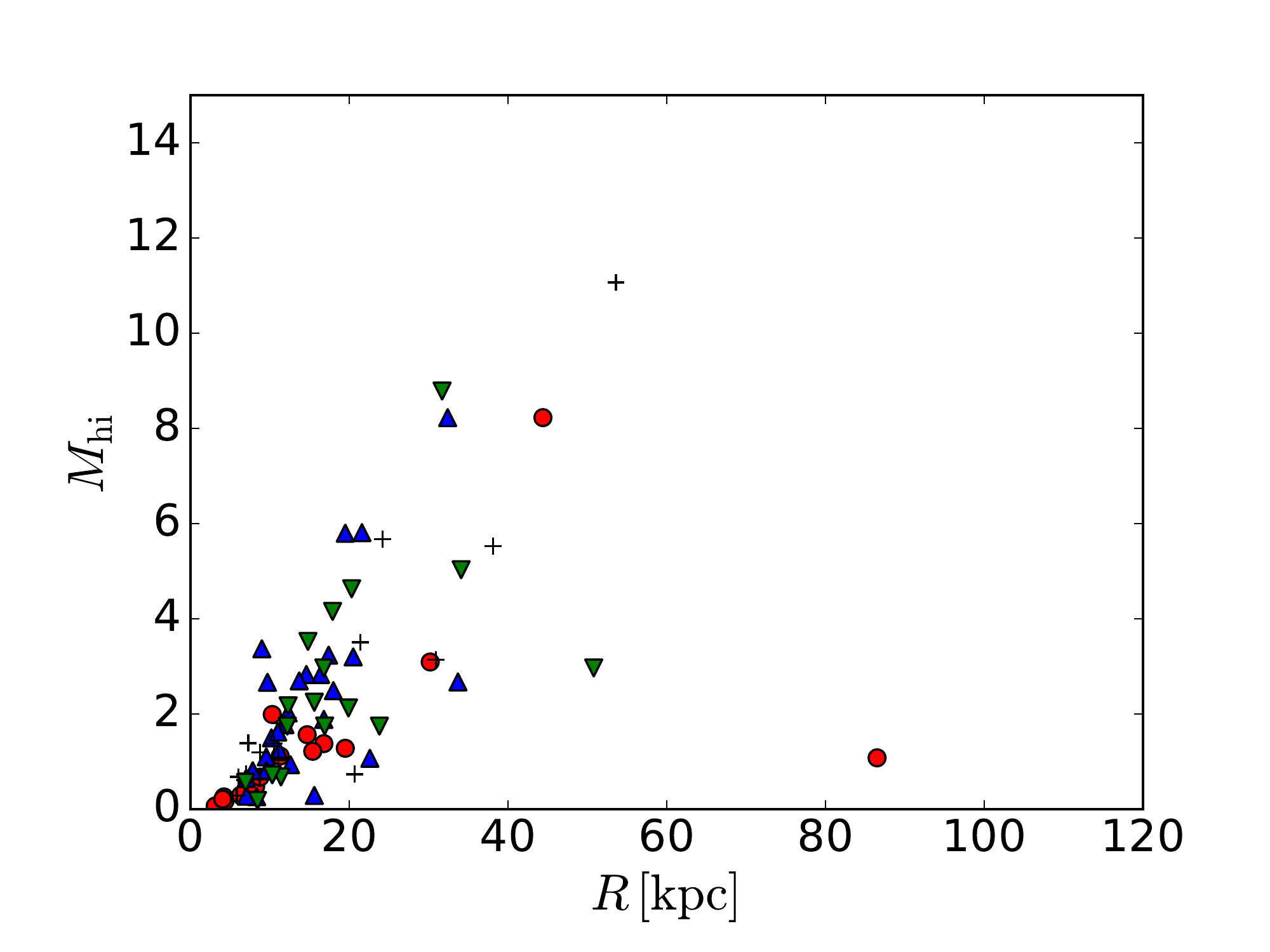}  \\
        \label{fig:plot_Rhi_vsR}         
        \includegraphics[width=0.48\textwidth]{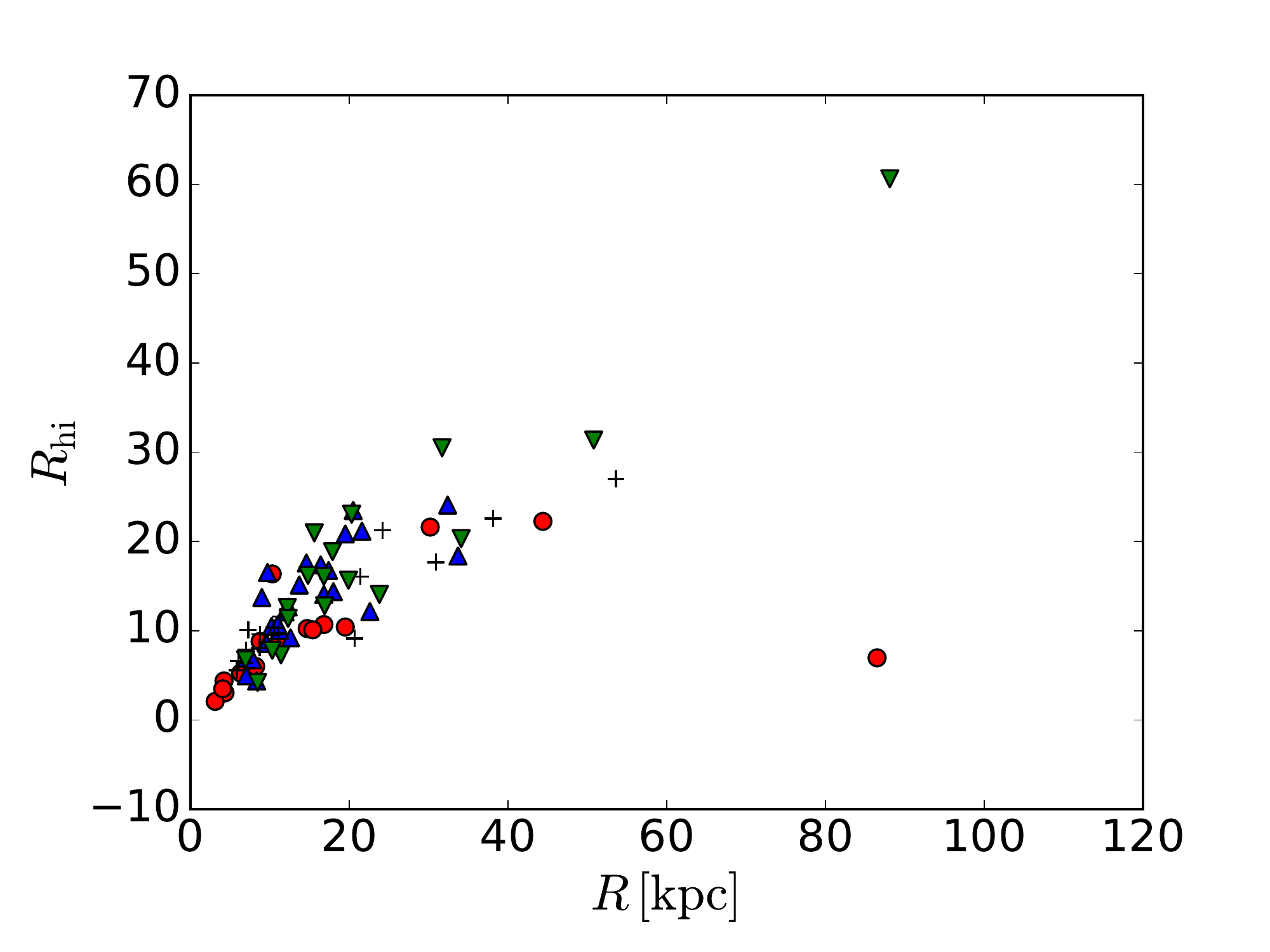}   
        \label{fig:plot_Vf_vsR}         
        \includegraphics[width=0.48\textwidth]{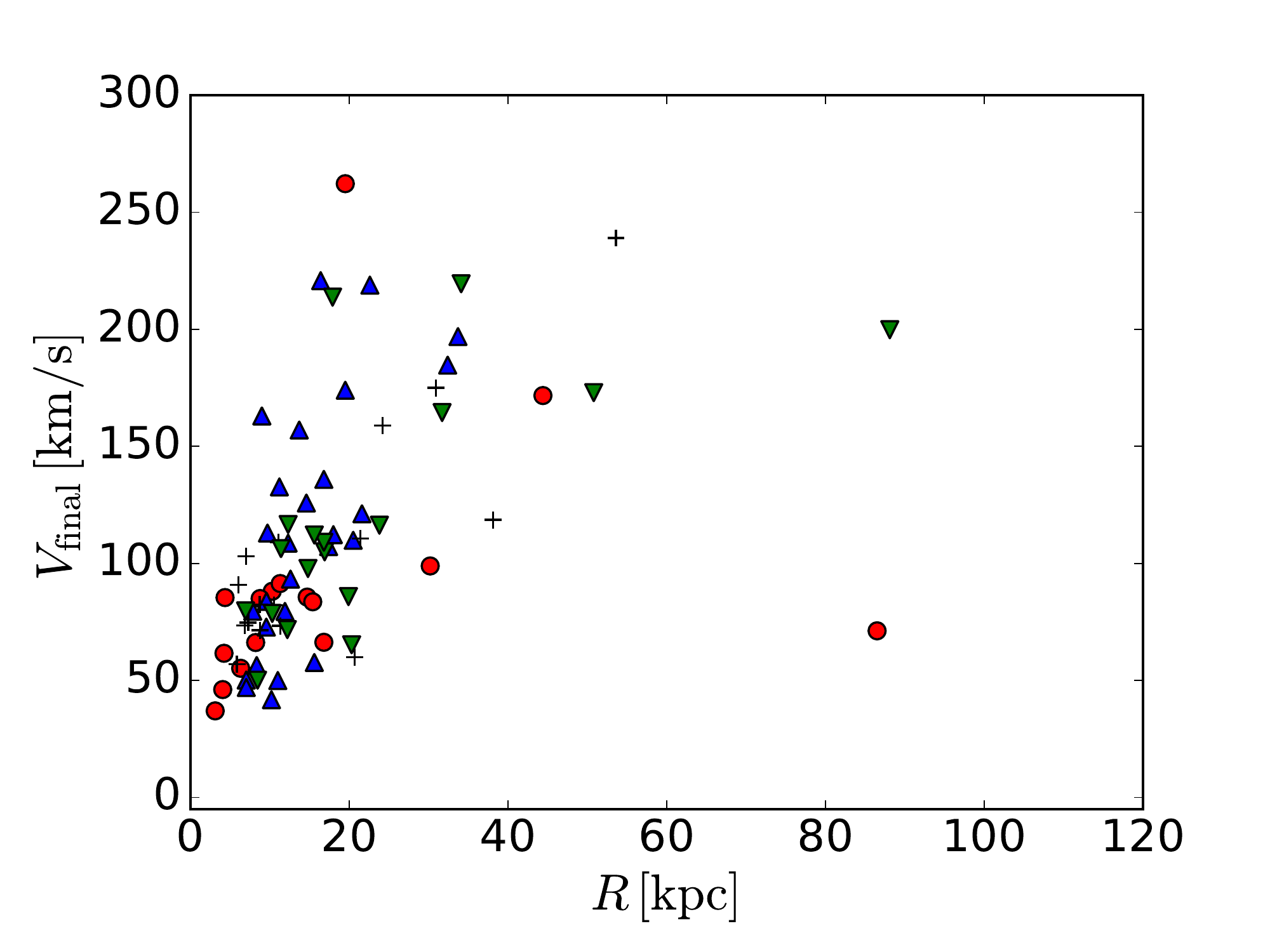}  
   \caption{Scattering of $R$ vs: f) $R_{{\rm d}}$, g) $M_{{\rm hi}}$, h) $R_{{\rm hi}}$, and i) $V_{{\rm final}}$. In all distributions, the number of multistates used to model the DM component is: One (red circles), two (blue up triangles), three (black crosses), and four (green down triangles).}
   \label{fig:VarVsR2}                
\end{figure*}

\begin{figure*}
   \centering
        \label{fig:plot_D_vsrho01}         
        \includegraphics[width=0.48\textwidth]{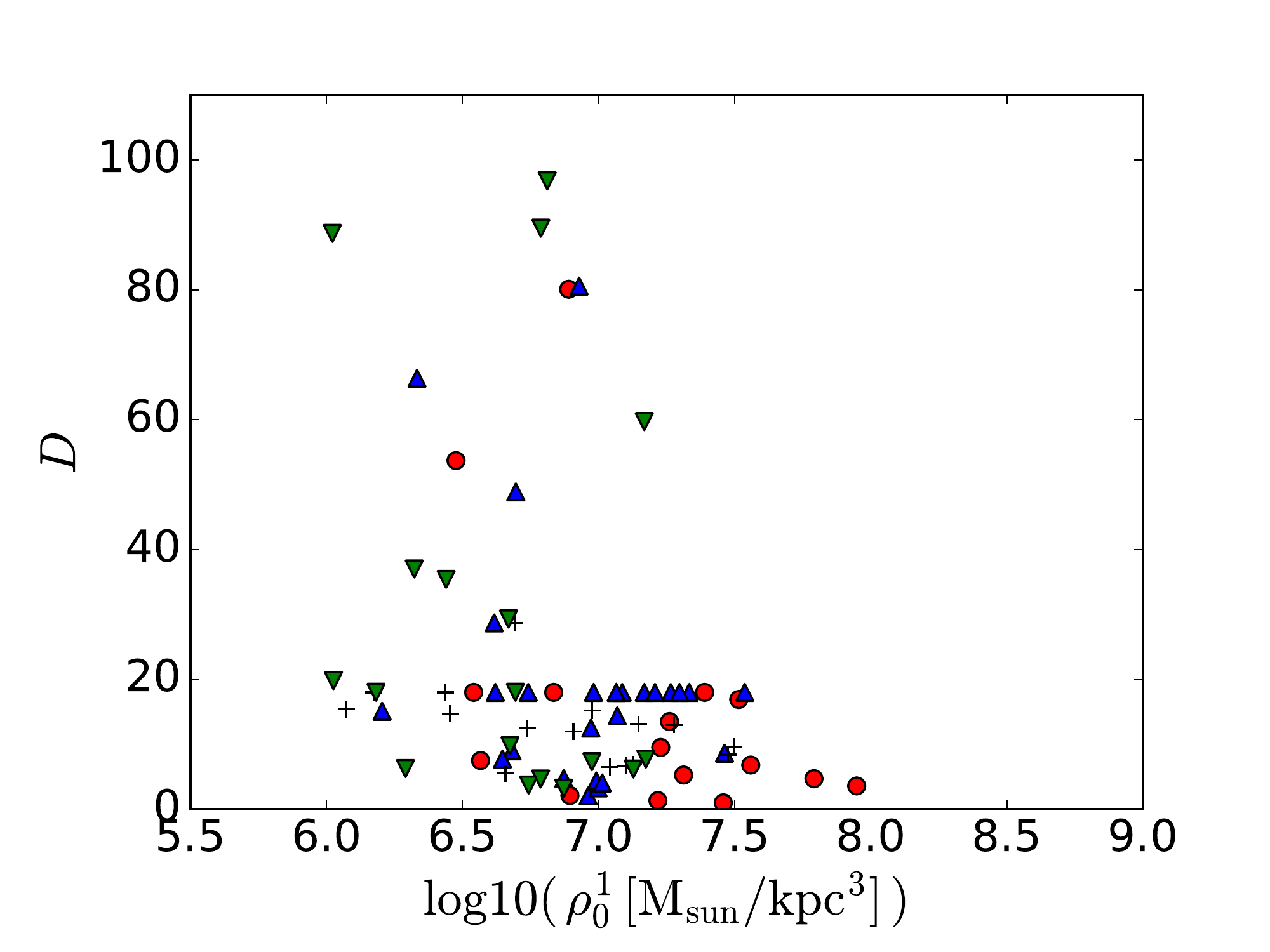} 
        \label{fig:plot_i_vsrho01}         
        \includegraphics[width=0.48\textwidth]{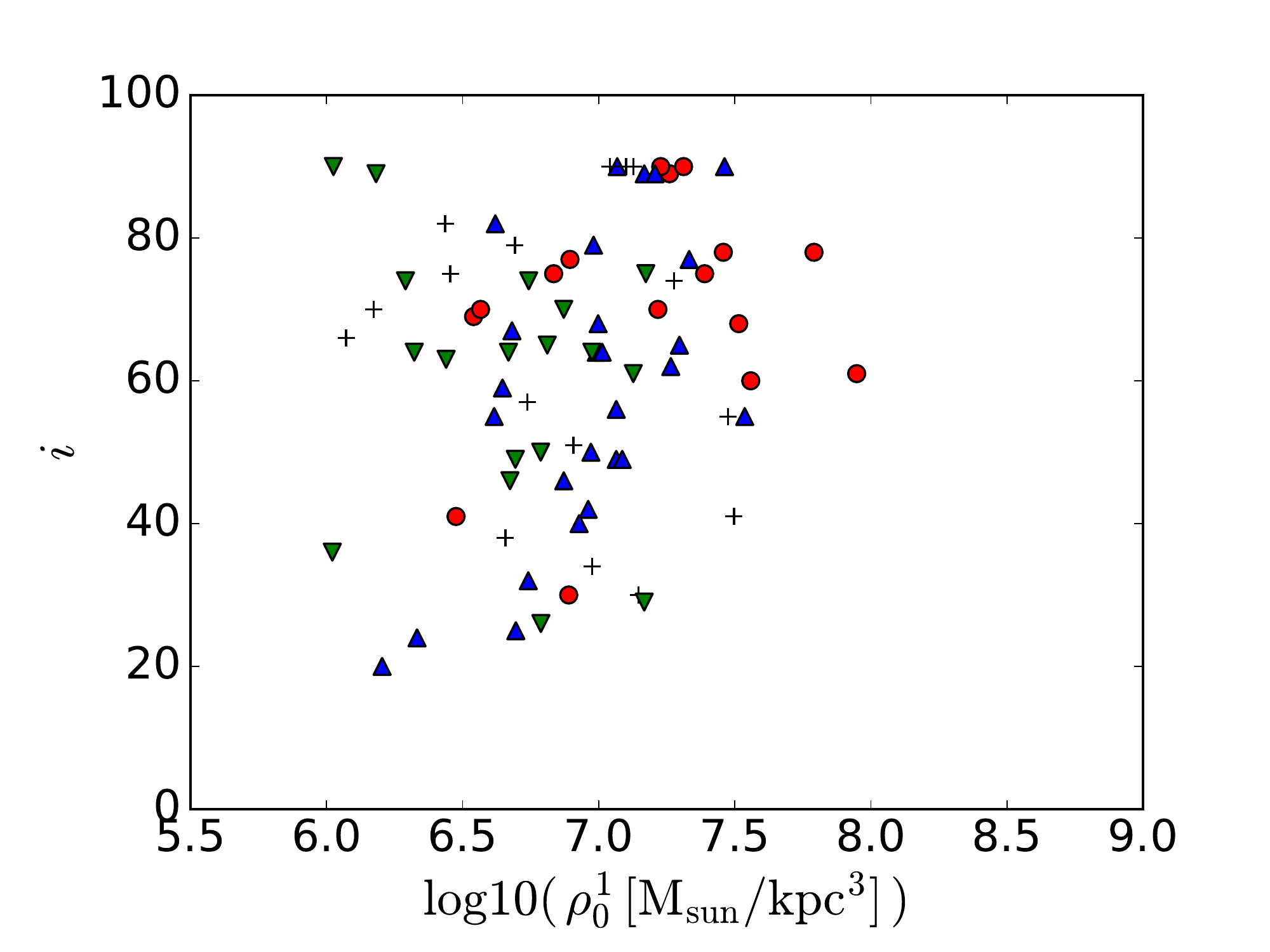}  \\
        \label{fig:plot_L_vsrho01}         
        \includegraphics[width=0.48\textwidth]{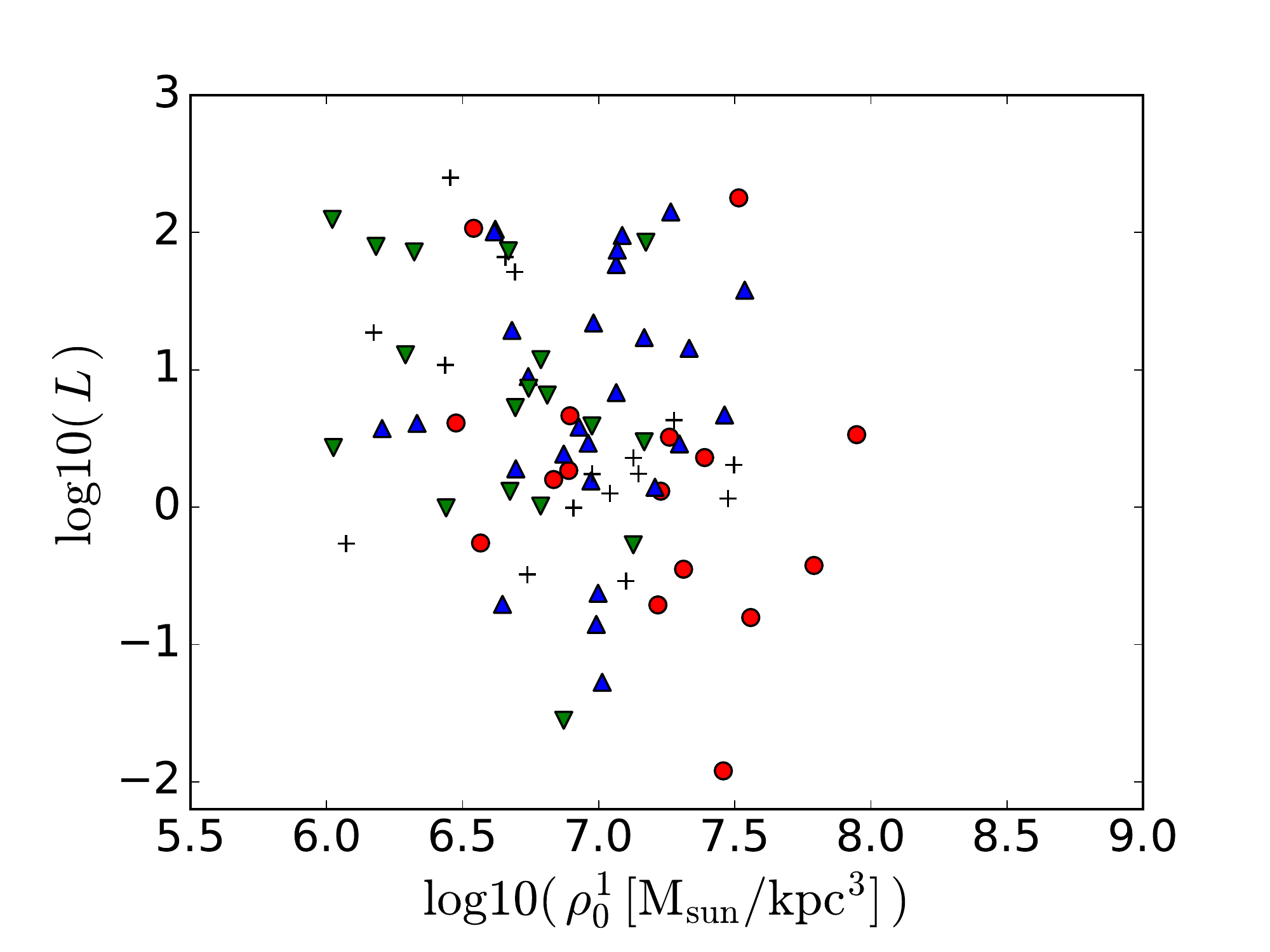}    
        \label{fig:plot_Reff_vsrho01}         
        \includegraphics[width=0.48\textwidth]{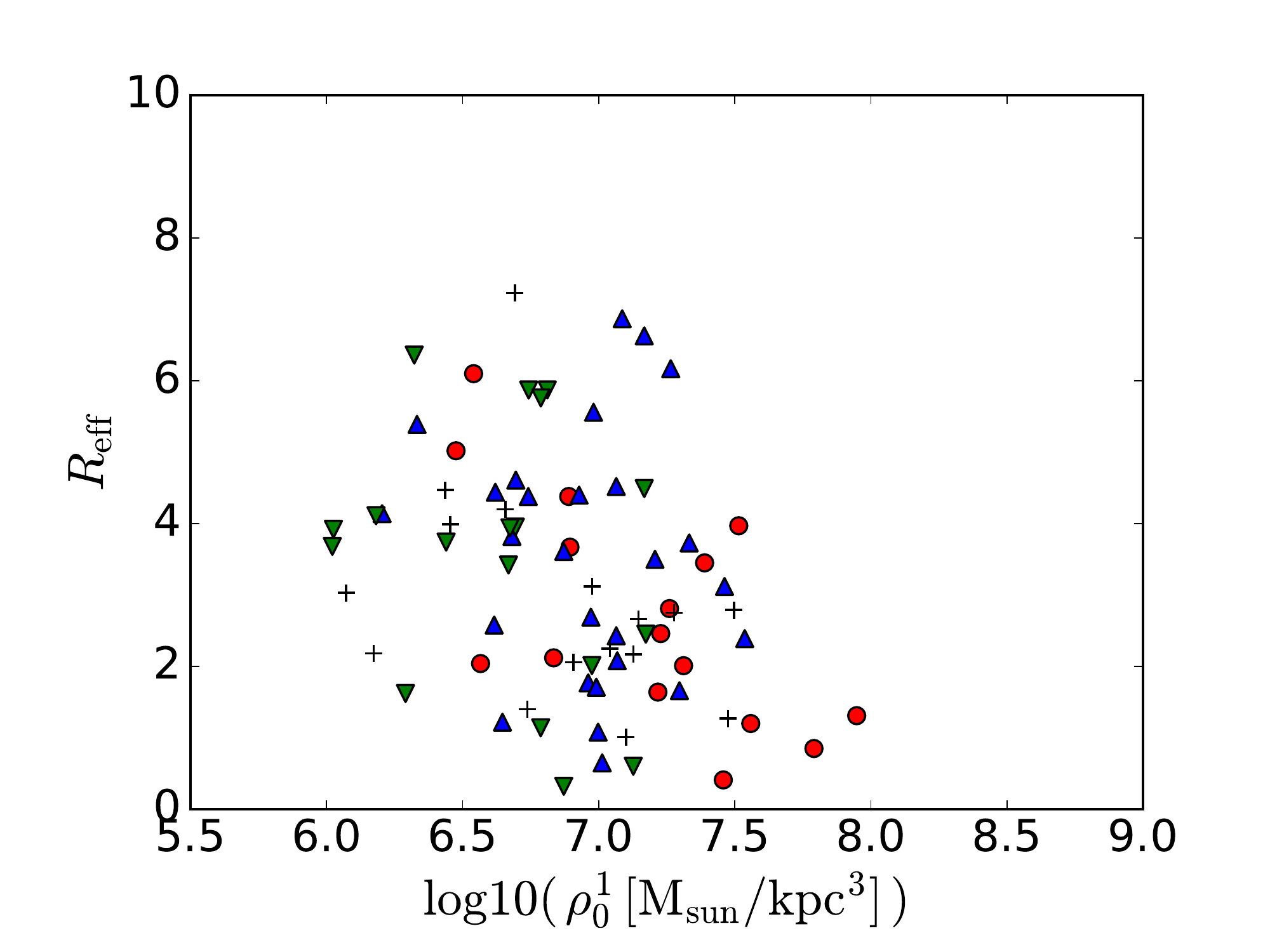}      
   \caption{Scattering of ground state $\rho_0^1$ vs: a) $D$, b) $i$, c) $\log10(L)$, d) $R_{{\rm eff}}$. In all distributions, the number of multistates used to model the DM component is: One (red circles), two (blue up triangles), three (black crosses), and four (green down triangles).}
   \label{fig:VarVsrho01_1}                
\end{figure*}

\begin{figure*}
   \centering
        \label{fig:plot_Rd_vsrho01}         
        \includegraphics[width=0.48\textwidth]{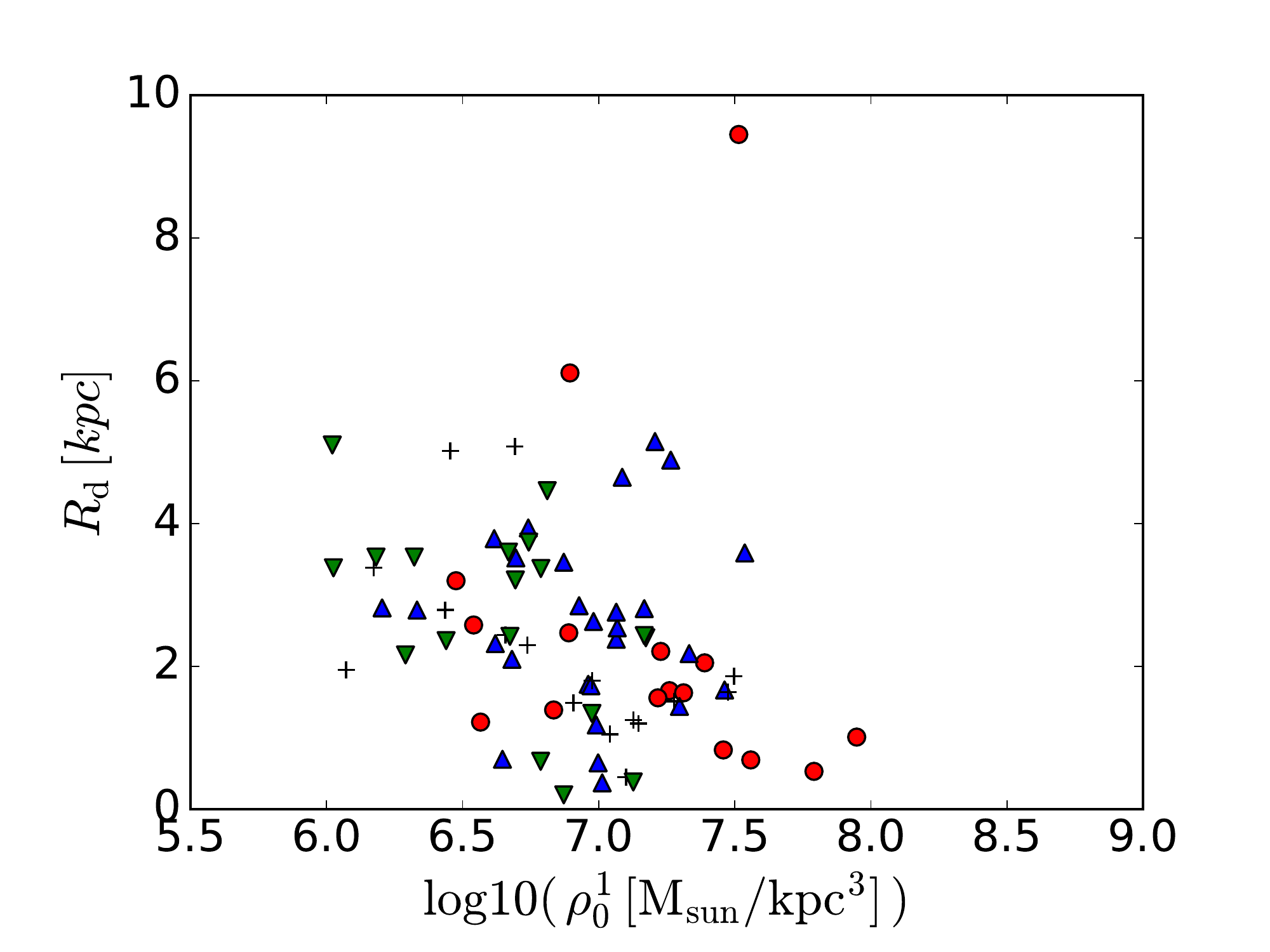}    
        \label{fig:plot_Mhi_vsrho01}         
        \includegraphics[width=0.48\textwidth]{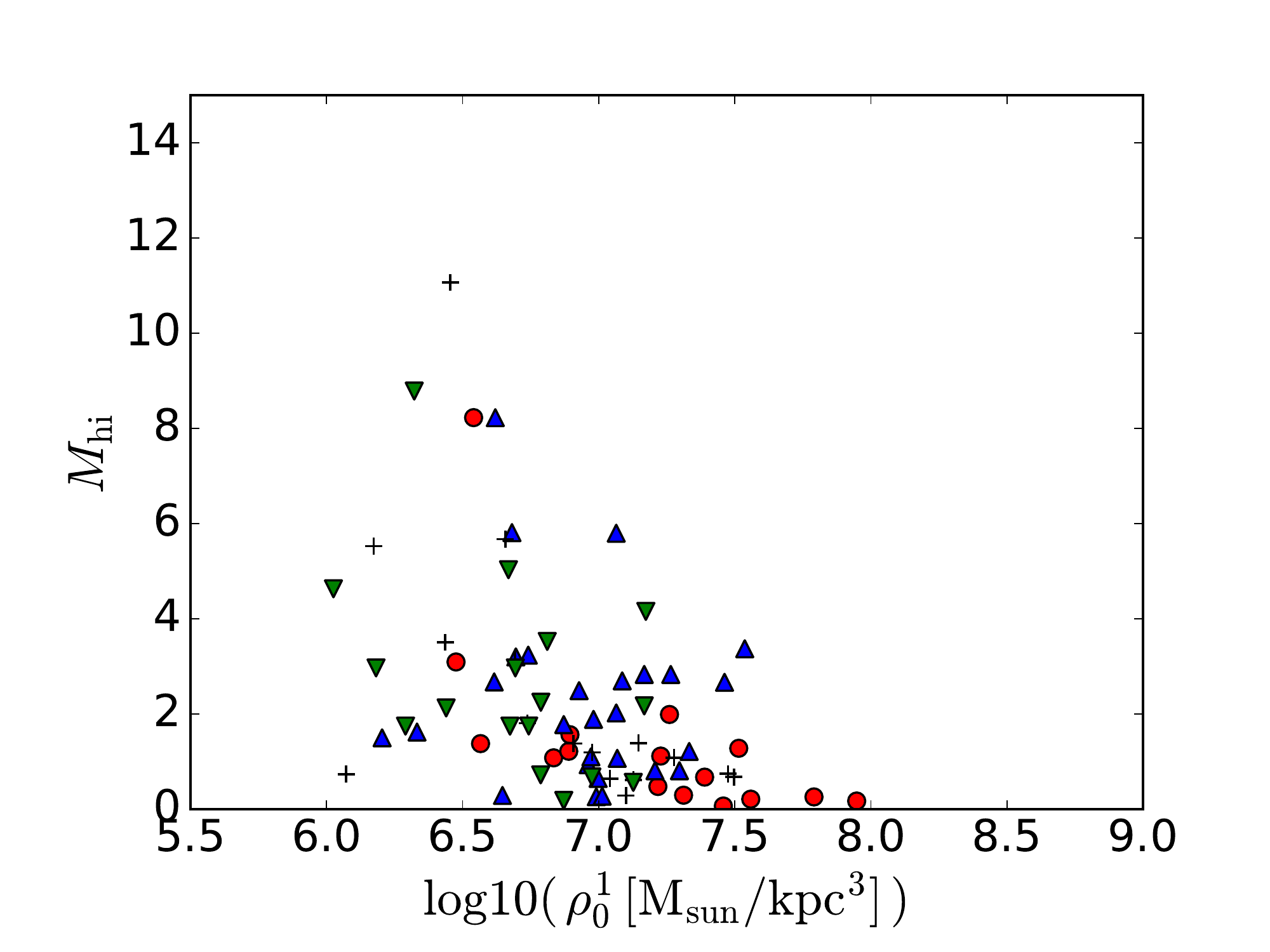}  \\
        \label{fig:plot_Rhi_vsrho01}         
        \includegraphics[width=0.48\textwidth]{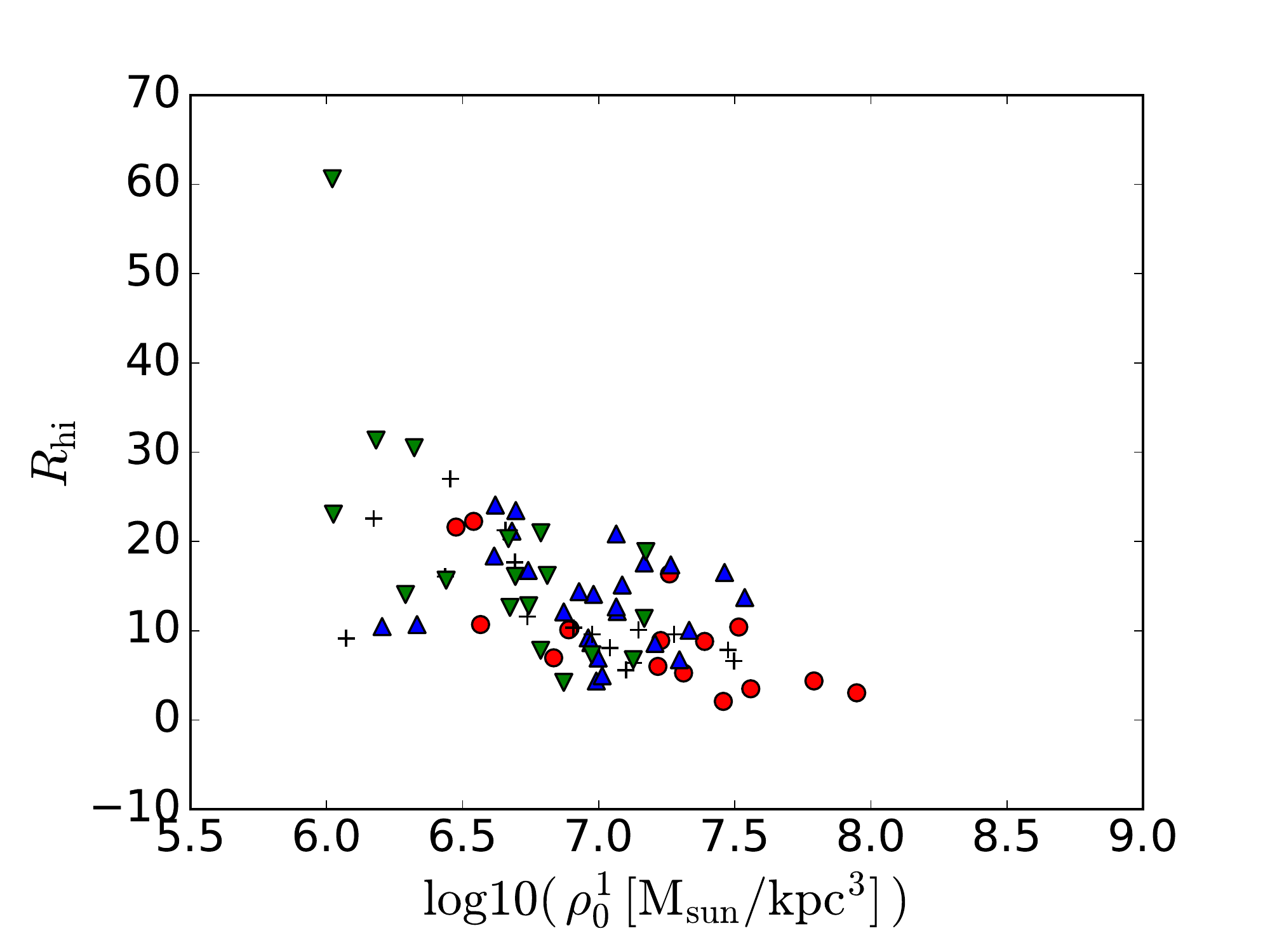}   
        \label{fig:plot_Vf_vsrho01}         
        \includegraphics[width=0.48\textwidth]{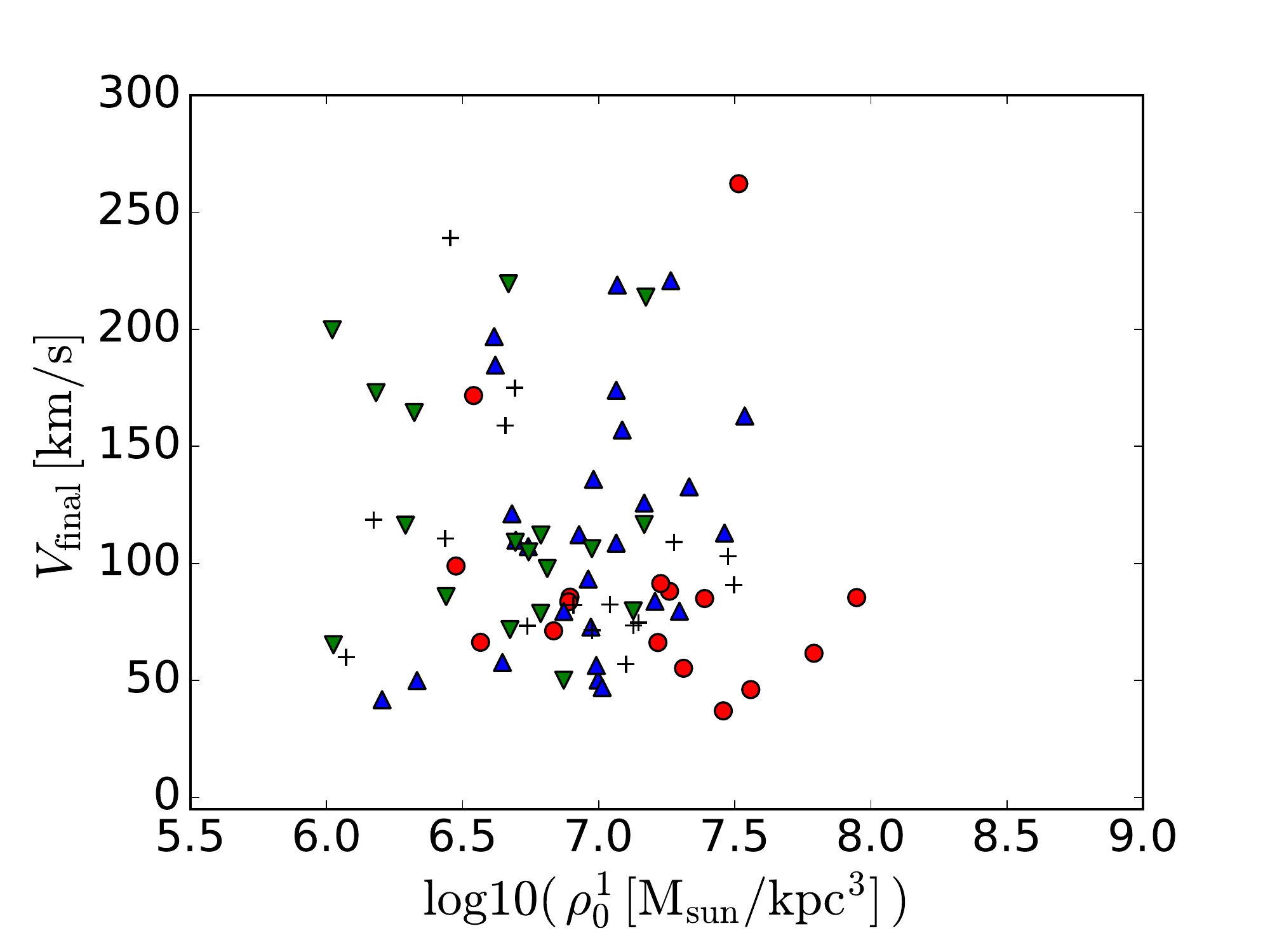}    
   \caption{Scattering of ground state $\rho_0^1$ vs: e) $\log10(S_{{\rm eff}})$, f) $R_{{\rm d}}$, g) $M_{{\rm hi}}$, h) $R_{{\rm hi}}$, and i) $V_{{\rm final}}$. In all distributions, the number of multistates used to model the DM component is: One (red circles), two (blue up triangles), three (black crosses), and four (green down triangles).}
   \label{fig:VarVsrho01_2}                
\end{figure*}

Finally, in Figs. \ref{fig:VarVsR1} and \ref{fig:VarVsR2} we show the scattering relation between the galactic properties shown in Tab. \ref{Tab:PropGalaxy} and $R$ classified according to the number of SFDM states. In an analogous way, it is presented in Figs. \ref{fig:VarVsrho01_1} and \ref{fig:VarVsrho01_2} their scattering with respect to ground state.

\begin{table}
\tbl{Correlations between Galaxy properties (left column) and SFDM model parameters (top row). Also, it is included a correlation between the parameters of the model. Notice that $\Upsilon_d=0.7$ and $\Upsilon_b=0.5$ are fixed on the fit.}
{\begin{tabular}{@{}ccccccc@{}} \toprule
Parameter&R&$\rho_0^1$&$\rho_0^2$&$\rho_0^3$&$\rho_0^4$&$\log_{10}$(M$_T$)\\ \colrule
                     $D$ &  $0.32$ & $-0.25$ & $-0.12$ & $-0.29$ & $-0.32$ & $0.30$ \\ 
                     $i$ &  $0.02$ & $0.15$  & $-0.03$ & $-0.12$ & $0.02$  & $0.05$ \\ 
                     $L$ &  $0.54$ & $-0.10$ & $-0.02$ & $-0.32$ & $0.04$  & $0.63$ \\ 
         $R_{{\rm eff}}$ &  $0.28$ & $-0.33$ & $-0.07$ & $-0.50$ & $-0.44$ & $0.45$ \\ 
    $\Sigma_{{\rm eff}}$ &  $0.43$ & $-0.03$ & $ 0.01$ & $-0.28$ & $0.42$  & $0.57$ \\ 
           $R_{{\rm d}}$ &  $0.37$ & $-0.19$ & $-0.06$ & $-0.56$ & $-0.47$ & $0.53$ \\ 
      $\Sigma_{{\rm d}}$ &  $0.25$ & $-0.09$ & $-0.06$ & $-0.14$ & $0.01$  & $0.31$ \\ 
          $M_{{\rm hi}}$ &  $0.72$ & $-0.31$ & $-0.23$ & $-0.39$ & $-0.28$ & $0.60$ \\ 
          $R_{{\rm hi}}$ &  $0.72$ & $-0.42$ & $-0.28$ & $-0.48$ & $-0.34$ & $0.68$ \\ 
       $V_{{\rm final}}$ &  $0.50$ & $-0.09$ & $0.13$  & $-0.33$ & $0.22$  & $0.70$ \\
\colrule
		             $R$ & $1.00$  &    $-$  & $-$     & $-$     & $-$     & $0.82$ \\    
          $\rho^{1}_{0}$ & $-0.39$ & $1.00$  & $-$     & $-$     & $-$     & $-0.35$ \\ 
          $\rho^{2}_{0}$ & $-0.35$ & $0.79$  & $1.00$  & $-$     & $-$     & $0.01$  \\ 
          $\rho^{3}_{0}$ & $-0.49$ & $0.78$  & $0.71$  & $1.00$  & $-$     & $-0.15$ \\ 
          $\rho^{4}_{0}$ & $-0.35$ & $0.75$  & $0.39$  & $0.62$  & $1.00$  & $0.11$  \\ 
\botrule
\end{tabular} \label{Tab:Corr}}
\end{table}

As a final comment, we compare the results of $\Upsilon_b$ and $\Upsilon_d$ fixed, with that they are floating. In general, we observe slightly fluctuation on the correlations, being the majority of the correlations preserved. It is also possible to notice that correlations close to limits between regions are sensitive to change. 

In this vein, in Table \ref{Tab:Corr_UdUb} are presented the results of the correlation when $\Upsilon_b$ and $\Upsilon_d$ are floating in the fit. The mean of the $\Upsilon_d$ and $\Upsilon_b$ best fit values are $0.52 \pm 0.11$ and $0.59 \pm 0.23$ respectively, where the uncertainties are the standard deviations.

\begin{table}
\tbl{Correlations between Galaxy properties (left column) and SFDM model parameters (top row) with $\Upsilon_d$ and $\Upsilon_b$ considered as a free parameter. Also, it is included a correlation between the parameters of the model. The mean value of $\Upsilon_d$ and $\Upsilon_b$ best fits are $0.52 \pm 0.11$ and $0.59 \pm 0.23$ respectively.}
{\begin{tabular}{@{}ccccccc@{}} \toprule
               Parameter &     $R$ &$\rho_0^1$&$\rho_0^2$&$\rho_0^3$&$\rho_0^4$&$\log_{10}$(M$_T$)\\ \hline
                     $D$ &  $0.22$ & $-0.27$ & $-0.15$ & $-0.28$ & $-0.22$ & $ 0.28$ \\ 
                     $i$ &  $0.05$ & $ 0.07$ & $-0.03$ & $-0.16$ & $-0.30$ & $-0.01$ \\ 
                     $L$ &  $0.52$ & $-0.17$ & $-0.23$ & $-0.18$ & $ 0.09$ & $ 0.57$ \\ 
         $R_{{\rm eff}}$ &  $0.31$ & $-0.34$ & $-0.18$ & $-0.40$ & $-0.26$ & $ 0.42$ \\ 
    $\Sigma_{{\rm eff}}$ &  $0.45$ & $-0.07$ & $-0.15$ & $-0.10$ & $ 0.19$ & $ 0.56$ \\ 
           $R_{{\rm d}}$ &  $0.45$ & $-0.30$ & $-0.21$ & $-0.48$ & $-0.37$ & $ 0.51$ \\ 
      $\Sigma_{{\rm d}}$ &  $0.29$ & $-0.14$ & $-0.18$ & $-0.17$ & $-0.02$ & $ 0.30$ \\ 
          $M_{{\rm hi}}$ &  $0.47$ & $-0.36$ & $-0.32$ & $-0.36$ & $-0.30$ & $ 0.44$ \\ 
          $R_{{\rm hi}}$ &  $0.51$ & $-0.52$ & $-0.43$ & $-0.52$ & $-0.44$ & $ 0.52$ \\ 
       $V_{{\rm final}}$ &  $0.60$ & $-0.19$ & $-0.20$ & $-0.20$ & $ 0.14$ & $ 0.76$ \\
\colrule
		             $R$ & $ 1.00$ &    $-$  & $-$     & $-$     & $-$     & $ 0.90$ \\    
          $\rho^{1}_{0}$ & $-0.27$ & $1.00$  & $-$     & $-$     & $-$     & $-0.24$ \\ 
          $\rho^{2}_{0}$ & $-0.47$ & $0.66$  & $1.00$  & $-$     & $-$     & $ 0.24$  \\ 
          $\rho^{3}_{0}$ & $-0.58$ & $0.86$  & $0.72$  & $1.00$  & $-$     & $-0.23$ \\ 
          $\rho^{4}_{0}$ & $-0.47$ & $0.86$  & $0.39$  & $0.89$  & $1.00$  & $ 0.01$  \\ 
\botrule
\end{tabular} \label{Tab:Corr_UdUb}}
\end{table}

\section{Conclusions and Outlooks} \label{Con}

We studied the multistate SFDM model for searching correlation between its parameters and galaxy properties. We focus our attention on the inclination parameter $i$ with the aim to establish a null correlation region, where we do not expect a natural relation with the DM halos. Indeed, as we observe in Tab. \ref{Tab:Corr}, our results show fluctuations around zero correlation. Based on this relation, it is presented in previous section a null correlation region in the interval $[-0.30,0.30]$; having the capability to decide if exist a correlation between different observation parameters.

The first correlation that catches our attention, is the relation between the size of galaxy halo and galaxy luminosity. When we consider galaxies up to $80\,$Mpc distance, we have a null correlation between these variables; being supported by the correlation between $D$ and halo mass. In addition, with higher statistics, we can give strong support to the fact that large galaxies are only observed at far away distances.

Other important features are the strong correlation with variables associated with the neutral hydrogen and the halo size, in this case we found that big size halos generate more presence of baryonic matter and in this particular case, more presence of neutral hydrogen.

In addition, our generalized results for the scalar field mass coincide with previous constrictions shown in the literature. So in general, associate more states to the study, generates a better fit to the value of the SFDM mass. Other observables that present low correlation, need more statistics, in order to confirm or refute the existence of such correlation. We also notice evidence of auto-interaction between the different SFDM states, suggesting that in future works it is necessary to consider crossed elements in the equation for the BEC and in the correlation analysis.

Finally, due to the low statistics, we remark that the results presented are sensitive to changes, being necessary further studies with higher statistics to have a more precise correlations and confirm if our correlations are conserved. However, this is work that will be presented elsewhere.

\section*{Acknowledgments}
We would like to thank the referee for thoughtful comments which helped to improve the manuscript. In addition, we acknowledge the enlightening conversation with Juan Maga\~na, providing us with important points about some features of the scalar field, also we appreciate the invaluable help of Professors Luis Ure\~na, Mario Rodr\'iguez and Jorge Mastache in the first draft reading together with their valuable comments and suggestions.
A.H. acknowledges SNI and M.A.G.-A. acknowledges support from CONACYT research fellow, SNI, and Instituto Avanzado de Cosmolog\'ia (IAC) collaborations.

\bibliographystyle{ws-ijmpd}
\bibliography{librero0}

\end{document}